\documentclass[aps,prb,twocolumn,floats,epsfig,pdflatex]{revtex4}
\usepackage{color,ulem,verbatim,bbold,float,wrapfig}
\usepackage{amssymb,amsbsy,amsmath,mathrsfs}
\usepackage{soul}
\usepackage{epsfig,subfigure}

\usepackage[colorlinks,linktocpage,bookmarks=false,citecolor=blue,linkcolor=red,urlcolor=blue]{hyperref}

\newcommand{\beq}{\begin{equation}}
\newcommand{\eeq}{\end{equation}}
\newcommand{\bea}{\begin{eqnarray}}
\newcommand{\eea}{\end{eqnarray}}

\begin{document}

\title{Analytic approaches to periodically driven closed quantum
systems: Methods and Applications}

\author{Arnab Sen$^{(1)}$, Diptiman Sen$^{(2)}$, and K. Sengupta$^{(1)}$}

\affiliation{$^{(1)}$School of Physical Sciences, Indian Association for the
Cultivation of Science, 2A and 2B Raja S. C. Mullick Road, Jadavpur 700032,
India \\
$^{(2)}$Center for High Energy Physics and Department of Physics,
Indian Institute of Science, Bengaluru 560012, India}

\date{\today}

\begin{abstract}

We present a brief overview of some of the analytic perturbative
techniques for the computation of the Floquet Hamiltonian for a
periodically driven, or Floquet, quantum many-body system. The key
technical points about each of the methods discussed are presented in a
pedagogical manner. They are followed by a brief account of some
chosen phenomena where these methods have provided useful insights.
We provide an extensive discussion of the Floquet-Magnus expansion,
the adiabatic-impulse approximation, and the Floquet perturbation
theory. This is followed by a relatively short discourse on the
rotating wave approximation, a Floquet-Magnus resummation technique
and the Hamiltonian flow method. We also provide a discussion of
some open problems which may possibly be addressed using these methods.

\end{abstract}

\maketitle

\section{Introduction}
\label{sec:int}

The physics of periodically driven, or Floquet, quantum many-body
systems has received tremendous attention in recent
times~\cite{rev1,rev2,rev3,rev4,rev5,rev6}. This is due to the fact
that such driven systems exhibit a gamut of interesting phenomena
which have no analogs either in equilibrium closed quantum systems
or in systems taken out of equilibrium using quench or ramp
protocols \cite{rev7, rev8}. Moreover, in recent times ultracold
atoms in optical lattices, trapped ions and superconducting qubits
have provided the much needed experimental platforms where
theoretical results involving such driven systems can be tested
\cite{rev9,exp1,exp2,exp3,exp4,exp5,revion,revqu}.

In periodically driven quantum many-body systems, where the
time-dependent Hamiltonian follows $H(t+nT)=H(t)$ for a fixed time
period $T=2\pi/\omega_D$ with $\omega_D$ being the associated
drive frequency and $n$ being an arbitrary integer, the
stroboscopic dynamics at times $t=nT$ is controlled by the Floquet
Hamiltonian $H_F$ \cite{flref1}. The Floquet Hamiltonian is a
Hermitian operator defined as the generator of the single-period
time-evolution operator, or the Floquet unitary $U(T)$, which equals
\bea U(T) &=& \mathcal{T}\left[ e^{-i\int_0^T dt H(t)/\hbar}\right] =
e^{ -i H_F T/\hbar} \label{HFdef}, \eea where $\mathcal{T}$ denotes
time-ordering. We note here that the time-ordering
makes it notoriously difficult to calculate $H_F$ for interacting
systems and one generally has to resort to various approximations.

Most of the phenomena in periodically driven closed quantum systems
which have attracted recent attention follow from the properties of
their corresponding Floquet Hamiltonians. For example, these
Hamiltonians in periodically driven systems may have topologically
non-trivial eigenstates even when the ground state of the
equilibrium system is topologically trivial
\cite{topo1,topo2,topo3,topo4,graphene1,topo5}. Thus the drive may
generate non-trivial topology which can be characterized by specific
topological invariants~\cite{topo5}. Such systems are also known to
exhibit dynamical transitions that arise from a change of the
properties of their Floquet Hamiltonian as a function of the drive
frequency; these transitions have no analog in quantum system in
equilibrium \cite{dtran1,dtran2}. Moreover, periodically driven
systems may lead to the realization of a time-crystalline state of
matter (a phase of matter which is disallowed in
equilibrium~\cite{tcnogo}); such a state (for a discrete time
crystal characterized by a $Z_n$ symmetry group) exhibits discrete
broken time translational symmetry so that its local correlation
functions develop $nT$-periodicity even when the Hamiltonian is
$T$-periodic \cite{qtc1,qtc2,qtc3}. Furthermore, such driven systems
exhibit dynamical freezing wherein the driven state of the system,
after $n$ periods of the drive at specific frequencies remains
arbitrarily closed to the initial state \cite{fr1,fr2,fr3,haldar19}. Also, it
is well-known that quantum systems in the presence of a periodic
drive may lead to dynamical localization where the drive leads to
suppression of the transport of particles in the system
\cite{dloc1,dloc1b,dloc2}. Finally, more recently, it was found that
there is a class of periodic drives which respects the conformal
symmetry of the underlying field theory; such driven conformal field
theories lead to drive-induced emergent spatial structures in the
energy density and correlation functions that have no analogs in
standard non-relativistic systems with external drive
\cite{fcft1,fcft2,fcft3}.

Another interesting feature of periodically driven quantum systems
can be understood from the perspective of the eigenstate
thermalization hypothesis (ETH) \cite{eth1,rev3}. It is generally
expected that all non-integrable ergodic quantum many-body systems
obey ETH in the thermodynamic limit. When driven periodically, such
systems absorb energy from the drive and heat up to an infinite
temperature steady state implying a featureless Floquet-ETH for
local correlation functions~\cite{LazaridesAM2014, PonteCPA2014}.
This has the interesting consequence of
the Floquet unitary $U(T)$ (Eq.~\eqref{HFdef}) resembling a random
matrix~\cite{AlessioR2014} with all its eigenstates mimicking random states
as far as local quantities are concerned, thus leading to a featureless
infinite temperature ensemble starting from all initial states.
However, the approach of the system to such a steady state, namely,
its prethermal behavior, in the presence of a periodic drive is not
well-understood and is the subject of many recent works
\cite{pretherm1, pretherm2,pretherm3, pretherm4}. It is generally
agreed upon that the time window $t_{\ast}$ for such a prethermal
regime depends on the drive frequency $t_{\ast} \sim e^{\hbar
\omega_D/J_{\rm{loc}}}$, where $J_{\rm{loc}}$ denotes a local energy scale,
in the high drive frequency limit \cite{pretherm1}. However, the extent of
this regime and the Floquet prethermalization mechanisms beyond high
frequencies in the intermediate or low drive frequency regime are
yet to be fully understood. This is a particularly relevant issue
since many ETH-violating phenomena in driven finite-sized systems
can be expected to occur for drive frequencies in the prethermal
regime in thermodynamically large systems, and such finite-sized
systems may be experimentally realized using various platforms like
ultracold atoms in optical lattices.

The violation of ETH in a thermodynamic many-body system may occur
due to loss of ergodicity due to the presence of a large number of
constants of motion as seen in integrable systems \cite{rev7} or due
to strong disorder as seen in the case of systems exhibiting many-body
localization \cite{rev10}. When such systems are periodically driven,
they reach steady states which can be qualitatively different from the standard
infinite temperature steady states of their ETH obeying counterparts
\cite{ss1, ss2, ss3}. Moreover, a wide range of quantum many-body
systems with constrained Hilbert spaces are known to host a special
class of many-body eigenstates called quantum scars in their Hilbert
space \cite{scar1,scar2,scar3, scar4}. It has been shown that the presence
of such quantum scars may change the quantum dynamics of such driven
systems \cite{scar1, scar4}; moreover, a periodic drive applied to such
systems with finite size may lead to reentrant transitions between
ergodic and non-ergodic behaviors as a function of the drive frequency
\cite{flscar1,flscar2}. This phenomenon, theoretically investigated
for a chain of Rydberg atoms, can be shown to follow from the
property of the Floquet Hamiltonian of the driven Rydberg chains
which can be experimentally realized using an ultracold atom setup
\cite{exp4}. Such finite chains have also been shown to exhibit both
dynamical freezing and novel ETH violating steady states \cite{fr3}.

Though these phenomena in periodically driven quantum systems follow from
the structure and properties of their Floquet Hamiltonian, the Floquet
Hamiltonian of a driven quantum system can, unfortunately, be computed
analytically for only a handful of cases. Therefore it is natural that
several approximate methods exist in the literature which try to obtain an
analytic, albeit perturbative, expression for $H_F$ (Eq.~\eqref{HFdef}).
These analytical results can then
be compared with exact numerical studies on finite-sized systems to
ascertain their accuracy and range of validity. In this review, we
will explore some of these methods with a pedagogical introduction
to the technical details for each followed by a short description of
a few chosen areas where these method have yielded useful results.
Three of these methods have been widely applied to a wide range of
driven systems and therefore deserve a somewhat long discourse.
These are the Floquet-Magnus expansion method (Sec.\ \ref{sec:FM}), the
adiabatic-impulse approximation (Sec.\ \ref{sec:adimp}), and the
Floquet perturbation theory (Sec.\ \ref{sec:fpt}). In addition, we
provide somewhat shorter discussions on the rotating wave
approximation, a recent Floquet-Magnus resummation technique, and
the Hamiltonian flow method in Sec.\ \ref{sec:om}. Finally, we
summarize this review and discuss a few open problems in the field
in Sec.\ \ref{sec:diss}.

\section{Floquet-Magnus expansion}
\label{sec:FM}

In this section, we will outline the calculation of $H_F$ in the
high driving frequency regime using a technique called
Floquet-Magnus (FM) expansion~\cite{Magnus, rev6} that formally
results in a series of the form \beq H_F = \sum_{n=0}^{\infty} ~T^n
\Omega_n. \label{FM} \eeq The FM expansion is the method of choice
to systematically calculate new terms in the Floquet Hamiltonian,
that may be otherwise difficult to generate in an equilibrium
setting, and thus manipulate out-of-equilibrium phases by
controlling the drive protocol.

The explicit expressions for the first three terms in Eq.~\eqref{FM}
are as follows:
\begin{eqnarray}
\Omega_0 &=& \frac{1}{T} \int_0^T dt_1 H(t_1), \label{FM3} \\
\Omega_1 &=& \frac{1}{2i\hbar T^2} \int_0^T dt_1 \int_0^{t_1} dt_2
[H(t_1), H(t_2)], \nonumber \\
\Omega_2 &=& -\frac{1}{6\hbar^2 T^3} \int_0^T dt_1 \int_0^{t_1} dt_2
\int_0^{t_2} dt_3 \nonumber \\
&&\left([H(t_1),[H(t_2),H(t_3)]]+[H(t_3),[H(t_2),H(t_1)]]\right).
\nonumber
\end{eqnarray}
The general expression for $\Omega_n$ (Eq.~\eqref{FM}) can be written
in terms of right-nested commutators of $H(t)$ as follows:
\begin{eqnarray}
&& \Omega_n = \frac{1}{(n+1)^2} \sum_{\sigma \in \mathcal{C}_{n+1}} (-1)^{d_b}
\frac{d_a ! d_b !}{n!} \label{FMn} \\
&& \times \frac{1}{i^n \hbar^n T^{n+1}} \int_0^{T} dt_1 \int_0^{t_1}
dt_2\cdots \int_0^{t_n} dt_{n+1} \nonumber \\
&& \times [H(t_{\sigma(1)}),[H(t_{\sigma(2)}),\cdots,[H(t_{\sigma(n)}),
H(t_{\sigma(n+1)})]\cdots]], \nonumber \end{eqnarray}
where $\sigma \in \mathcal{C}_{n+1}$ denotes a permutation of
$\{1,2,\cdots,n+1\}$ (sum is over the $(n+1)!$ permutations of
$\mathcal{C}_{n+1}$), $d_a$ ($d_b$) is the number of ascents
(descents) in the permutation $\sigma$ where $\sigma$ has an ascent
(a descent) in $i$ if $\sigma(i) < \sigma(i+1)$ ($\sigma(i) >
\sigma(i+1)$), $i=1,\cdots,n$ for $(i_1 i_2 \cdots
i_{n+1})=(\sigma(1) \sigma(2) \cdots \sigma(n+1))$, thus giving
$d_a+d_b=n$ for any permutation $\sigma$.

We will now indicate the essential steps required for the derivation of the
FM expansion (for more details, we refer the reader to Refs.~\onlinecite{rev6}
and \onlinecite{ArnalCC2018}). From standard quantum mechanics, the
propagator $U(t,t_0)$ defined by \beq |\psi(t) \rangle =
U(t,t_0) |\psi(t_0) \rangle , \mbox{~~~} \mathrm{where}
\mbox{~~~} U(t_0,t_0)=I \label{Ueq} \eeq (here $I$ is the
identity matrix and $|\psi(t)\rangle$ is the many-body wave function
of the system at time $t$), can be expressed in terms of the Dyson
series as follows:
\begin{eqnarray}
&&U(t,t_0) = I+ \sum_{n=1}^\infty P_n(t,t_0), \mbox{~~~}
\mathrm{where}\label{Dyson} \\
&&P_n(t,t_0)=\left( \frac{-i}{\hbar}\right)^n \int_{t_0}^t dt_1 \cdots
\int_{t_0}^{t_{n-1}}dt_n H(t_1) \cdots H(t_n). \nonumber
\end{eqnarray}
Since $U(T)=U(T,0)$ from Eq.~\eqref{HFdef}, we can simply
put $t_0=0$ and $t=T$ in Eq.~\eqref{Dyson} to obtain the
Dyson series for $U(T)$. Note that truncating the Dyson series does
not result in a unitary approximation for $U(T)$. From
Eqs.~\eqref{HFdef} and \eqref{Dyson}, it follows that \beq H_F =
\frac{i\hbar}{T} \ln \left( I+ \sum_{n=1}^\infty P_n
\right), \label{HFdyson} \eeq where we denote $P_n(T,0)$ by $P_n$ for brevity.
Using the series expansion for the logarithm in the above expression
(Eq.~\eqref{HFdyson}), expressing the LHS using Eq.~\eqref{FM}
and finally, matching terms with the same
powers of $H(t)$ allows one to express $\Omega_n$ (Eq.~\eqref{FMn})
in terms of $P_n$ (Eq.~\eqref{Dyson}). In particular, for the first
few terms, we get
\begin{eqnarray} \Omega_0 &=& \frac{i\hbar}{T}P_1, \nonumber \\
\Omega_1 &=& \frac{i \hbar}{T^2} \left(P_2-\frac{1}{2}P_1^2 \right),
\nonumber \\
\Omega_2 &=& \frac{i\hbar}{T^3} \left(P_3-\frac{1}{2}(P_1P_2+P_2P_1)
+ \frac{1}{3}P_1^3 \right), \label{magnus1} \end{eqnarray}
To express the RHS of $\Omega_1$ and $\Omega_2$ (Eq.~\eqref{magnus1})
in terms of right-nested commutators, we introduce the following
notation:
\begin{eqnarray} p(i_1 i_2 \cdots i_n)&=& \int_0^T dt_1 \int_0^{t_1}dt_2
\cdots \int_0^{t_{n-1}}dt_n \times \nonumber \\
&&H(t_{i_1})H(t_{i_2})\cdots H(t_{i_n}). \label{Bnotation} \end{eqnarray}
Using Fubini's theorem, \beq \int_0^a dy \int_y^a F(x,y) dx =
\int_0^a dx \int_0^x F(x,y) dy \label{Fubini}, \eeq it can then be
shown that
\begin{eqnarray} p(1)\cdot p(1) &=& p(12)+p(21), \nonumber \\
p(1)\cdot p(12) &=& p(123)+p(213)+p(312),\nonumber \\
p(12) \cdot p(1) &=& p(123)+p(132)+p(231), \nonumber \\
p(1) \cdot p(1) \cdot p(1) &=& p(123)+p(132)+p(213), \nonumber \\
&& + p(231)+p(312)+p(321). \label{permute} \end{eqnarray}
Using Eq.~\eqref{permute} in Eq.~\eqref{magnus1} gives Eq.~\eqref{FM3}.
For example, $p(12)-(1/2)(p(1)\cdot p(1))=(1/2)(p(12)-p(21))$ from
which the expression for $\Omega_1$ follows straightforwardly. It
should be noted that the RHS of Eq.~\eqref{permute} contains all
possible permutations of time ordering that are consistent with the
time ordering within the factors of the original products on the
LHS. For example, in the second line of Eq.~\eqref{permute}, terms
such as $p(132),p(231),p(321)$ do not appear because they are
inconsistent with the time ordering implied by the LHS $p(1) \cdot
p(12)$ that the second index is less than the third index while the
first index is arbitrary in $p(i_1 i_2 i_3)$. This structure
generalizes to higher orders as well allowing for the derivation of
$\Omega_n$ in terms of right-nested commutators (Eq.~\eqref{FMn}).

An important case where the integrals in Eq.~\eqref{FMn} may be
analytically computed is for a step-like drive between Hamiltonians
$H_1$ for duration $T_1$ and $H_2$ for duration $T_2$ where
$T=T_1+T_2$. Eq.~\eqref{FM} then reduces to the
Baker-Campbell-Hausdorff (BCH) formula where
\begin{eqnarray} Z &=& \ln \left(\exp(X) \exp(Y) \right) \label{BCH} \\
&=& X + Y + \frac{1}{2}[X,Y]+\frac{1}{12}[X-Y,[X,Y]]+\cdots, \nonumber
\end{eqnarray}
with the identification that $X = -iH_1T_1/\hbar$,
$Y=-iH_2T_2/\hbar$ and $Z = -iH_FT/\hbar$. In this section, we henceforth set
$\hbar=1$ for notational convenience.

We now summarize a few general points regarding the FM expansion
focussing on many-body lattice models with short-ranged Hamiltonians
and bounded local Hilbert spaces~\cite{rev5, rev1, pretherm1}. From
Eqs.~\eqref{FM3} and \eqref{FMn}, it is clear that while only
$\Omega_0 \neq 0$ if $[H(t), H(t')]= 0$ for $t \neq t'$; in the case
where $[H(t), H(t')]\neq 0$, the FM expansion (Eq.~\eqref{FM}) is an
infinite series in general. A sufficient (but not necessary)
condition for this infinite series to converge is that \beq
\frac{1}{\hbar} \int_0^T dt \|H(t) \|_2 < \pi, \label{convergeFM}
\eeq where $\| A \|_2$ denotes the spectral norm of a matrix $A$
that equals the square root of the largest eigenvalue of the matrix
$A^\dag A$. For short-ranged Hamiltonians, given that the energy is
extensive, we expect that $(1/\hbar) \int_0^T dt \|H(t) \|_2 \propto
N$ where $N$ denotes the number of degrees of freedom, which implies
that in general, Eq.~\eqref{convergeFM} cannot be satisfied for any
{\it finite} $T$ in a thermodynamically large system. In fact, the
weight of evidence suggests that the FM expansion is indeed
divergent for periodically driven interacting
systems~\cite{pretherm1} which eventually heat up to a featureless
infinite temperature ensemble at late times due to the absence of
energy conservation under driving~\cite{LazaridesAM2014,
PonteCPA2014}. Assuming that the Hamiltonian has at most $k$-local
terms (e.g., $k$-spin interactions in a quantum spin model on a
lattice), the higher-order terms in the expansion generate
progressively longer-ranged terms where $\Omega_n$ contains at most
$nk$-local terms. Thus, taking the infinite series for the
Floquet-Magnus expansion should amount to generating a $H_F$ that
resembles a random matrix~\cite{AlessioR2014} and hence mimics an
infinite temperature ensemble locally.

However, an important simplification happens at large drive
frequencies~\cite{pretherm1} which makes truncating this divergent
FM expansion up to a finite order physically meaningful. When the
drive frequency $\omega_D \gg J_{\mathrm loc}$ where $J_{\mathrm
loc}$ denotes the energy scale associated with local rearrangements
of the degrees of freedom in an interacting problem (which can be
deciphered from $H(t)$), there appears a large transient time $t_{*}
\sim \exp (\omega_D/J_{\mathrm loc})$ below which the system is in a
prethermal Floquet state that can be well described by a truncated
Floquet Hamiltonian $H_F^{(n)} = \sum_{m=0}^n T^m \Omega_m$ choosing
an optimum $n=n_0$. The heating is prevented in the prethermal
regime ($t \lesssim t_{*}$) because $H_F^{(n)}$ appears as a
conserved quantity at stroboscopic times, i.e., at times $t=nT$.
Moreover, and very importantly, the dynamics of local observables
can also be accurately described~\cite{pretherm1} by the unitary
dynamics generated from the truncated Floquet Hamiltonian
$H_F^{(n)}$ when $t \lesssim t_{*}$. For times $t \gg t_{*}$, the
system eventually flows to an infinite temperature ensemble.
Physically, a many-body system requires
$\mathrm{O}(\omega_D/J_{\mathrm{loc}}) \gg 1$ correlated local
rearrangements to absorb a single quantum of energy from the drive
when the drive frequency is large, hence implying a heating time
that scales as $\exp(\omega_D/J_{\mathrm{loc}})$.

We now give an example to show that non-trivial terms can be
generated in the FM expansion even at low orders ($\Omega_1$, etc in
Eq.~\eqref{FM3}) which may be otherwise difficult to generate in
static settings. To this end, we consider a model of spinless
fermions on a one-dimensional lattice where the Floquet driving is
chosen in such a manner that the problem is {{\it dynamically
localized}} without interactions. We then consider the interacting
problem and use the FM expansion to calculate the first few terms of
the Floquet Hamiltonian. Since the problem has no kinetic energy in
the Floquet Hamiltonian by construction (due to the dynamical
localization), these terms are entirely determined by the
interaction energy scale and the driving period $T$, and generate
density-dependent hoppings of the fermions as we show below
~\cite{dloc1b}.

To this end, let us consider the Hamiltonian
\begin{eqnarray}
H &=& H_{NI}+H_{I} \nonumber \\
&=& -\gamma \sum_{j=1}^N (c_j^\dagger c_{j+1} + \mathrm{H.c.}) + V
\sum_{j=1}^N n_j n_{j+1}, \label{fermion1} \end{eqnarray}
where $n_j=c_j^\dagger c_j$, $c_{N+1}=c_1$, and the number of sites, $N$,
is even. The Floquet structure is induced by a
periodic kicking Hamiltonian of the form \beq H_K =
\sum_{n=-\infty}^\infty \delta(t-nT) (\alpha N_e -\beta N_o),
\label{fermion2} \eeq where $N_e = \sum_{i \in e}n_i$ and $N_o =
\sum_{i \in o}n_i$ are the total number of fermions on the even and
odd sites, respectively. Considering the non-interacting limit of
$V=0$, and using the following special case of the BCH formula
(Eq.~\eqref{BCH}) when $[X,Y]=\zeta Y$,
\begin{equation}
\exp(X) \exp(Y) = \exp(\exp(\zeta)Y) \exp(X). \label{BCH_specialcase}
\end{equation}
Since $[n_j,c_j]=-c_j$ and $[n_j,c_j^\dagger] = c_j^\dagger$, we obtain
\begin{eqnarray} U(T) &=& U_K U_{NI} \nonumber \\
&=& \exp(-i(\alpha N_e -\beta N_o)) \exp(-iH_{NI}T) \nonumber \\
&=& \exp(+iH_{NI}T) \exp(-i(\alpha N_e -\beta N_o)), \label{algebra1}
\end{eqnarray}
when $\alpha+\beta=\pi$. Restricting to $\alpha=\pi$, $\beta=0$ so
that the periodic kicks (Eq.~\eqref{fermion2}) are applied only to the
even sites implies that \beq U(2T)=U^2(T) = \exp(-i2\pi
N_t)=I, \label{freezing} \eeq where $N_t$ is the total
number of fermions in the system. Thus, the non-interacting system is
strictly localized at intervals of $2T$. We now turn on $H_I$
(Eq.~\eqref{fermion1}) and compute $H_F$ as an expansion in powers of
the drive period $T$ (Eq.~\eqref{FM}). Since $H_I$ commutes with
$H_K$, it can be seen that
\begin{eqnarray} U^2(T) &=& \exp(-iH_F2T) \label{intfermions1} \\
&=& \exp(-i(-H_{NI}+H_I)T)\exp(-i(H_{NI}+H_I)T). \nonumber
\end{eqnarray}
We can now use the BCH formula (Eq.~\eqref{BCH}) to arrive at
\beq H_F = H_I +
\frac{iT}{2}[H_{NI},H_I]-\frac{T^2}{6}[H_{NI},[H_{NI},H_I]]+\cdots,
\label{intfermions2} \eeq which finally gives
\begin{widetext}
\begin{eqnarray}
H_F &=& V \sum_j n_j n_{j+1}-\frac{i\gamma TV}{2} \sum_j (c^\dagger_{j+1}
c_j -c^\dagger_j c_{j+1})(n_{j-1}-n_{j+2})-\frac{\gamma^2T^2 V}{3} \sum_j
\left(\frac{}{} (n_j-n_{j+1})(n_{j-1}-n_{j+2})\right. \nonumber \\
&+& \left. \frac{1}{2} (c_{j-1}^\dagger c_{j+1}+c_{j+1}^\dagger c_{j-1})
(n_{j+2}+n_{j-2}-2n_j) - (c^\dagger_{j-2}c_{j-1}-c^\dagger_{j-1}c_{j-2})
(c^\dagger_{j}c_{j+1}-c^\dagger_{j+1}c_{j})\right). \label{intfermions3}
\end{eqnarray}
\end{widetext}
Thus, the Floquet Hamiltonian (Eq.~\eqref{intfermions3}) contains
density-dependent fermion hoppings and pairwise hoppings apart from the
usual density-density interactions induced by $H_I$ (Eq.~\eqref{fermion1}).

Before concluding this section, we briefly discuss another
incarnation of Floquet prethermalization that allows the realization
of prethermal versions of nonequilibrium phases like Floquet time
crystals~\cite{qtc1,qtc2,qtc3}, but without the necessity of strong
disorder~\cite{ElseBN2017}. Such a prethermalization occurs when the
drive frequency $\omega_D$ is greater than {\it all but one} of the
local scales of the Hamiltonian. Let the time-dependent Hamiltonian
$H(t)$ be of the form \beq H(t)=H_0(t)+V(t), \label{ptc1} \eeq where
both $H_0(t)$ and $V(t)$ are periodic functions with period $T$.
Furthermore, $\lambda T \ll 1$ where $\lambda$ is the local energy
scale of $V(t)$. Importantly, the term in the Hamiltonian with the
large coupling (comparable to the drive frequency $\omega_D$) needs
to take a special form to avoid rapid heating. $H_0(t)$ has the
property that it generates a trivial time evolution over $M$ time
cycles, i.e.,
\begin{eqnarray} X^M &=& I, \mbox{~~~~~} \mathrm{where} \nonumber \\
X &=& \mathcal{T} \exp \left(-i \int_0^T dt H_0(t) \right).
\label{ptc2} \end{eqnarray}
Going to the interaction picture (where $V(t)$ is the
``interaction'' term), we see that \beq U(MT) = \mathcal{T}\exp
\left(-i\int_0^{MT} dt V_{int}(t) \right), \label{ptc3} \eeq where
$V_{int}(t) = U_0(t,0)^\dagger V(t) U_0(t,0)$
with $U_0$ being the propagator for $H_0(t)$. Since
$U_0(MT)=X^M= I$, the time evolution operator $U(MT)$ is
identical in the interaction and Schr\"odinger pictures.
Rescaling time as $t \rightarrow t/\lambda$, Eq.~\eqref{ptc3} then
describes a system being periodically driven at a large frequency
$\tilde{\omega}_D = 2\pi/(\lambda MT)$ by a drive of local strength
$1$ where standard Floquet prethermalization results
apply~\cite{pretherm1} for $\tilde{\omega}_D \gg 1$ resulting
in a large prethermal time $t_* \sim \exp(\tilde{\omega}_D)$.

Generalizing the ideas in Ref.~\onlinecite{AbaninRHH2017},
Ref.~\onlinecite{ElseBN2017} showed that within the prethermal
window, the Floquet unitary $U(T)$ can be well approximated by \beq
U(T) \approx \mathcal{V}^\dagger \left(X
\exp(-iDT)\right)\mathcal{V}, \label{ptc4} \eeq where $\mathcal{V}$
is a time-independent local unitary rotation, and $D$ is a local
Hamiltonian that has the property $[D,X]=0$. Hence the stroboscopic
time evolution has an {\it emergent} symmetry
$\mathcal{V}X\mathcal{V}^\dagger$ that commutes with $U(T)$ even
though $H(t)$ has no such symmetry. Interesting prethermal phases
can be stabilized when $X$ can be interpreted as a symmetry
that can be spontaneously broken due to the choice of the initial state
and the dimensionality of the system.

For example, to stabilize a prethermal Floquet time crystal, an
Ising ferromagnet can be considered on the square lattice with a
longitudinal field applied to break the Ising symmetry explicitly,
and a time-dependent transverse field providing the periodic
driving~\cite{ElseBN2017}. Thus,
\begin{eqnarray} H_0(t) &=& -\sum_i h_x(t) \sigma_i^x, \nonumber \\
V &=& -J \sum_{\langle ij \rangle} \sigma_i^z \sigma_j^z - h_z
\sum_i \sigma_i^z. \label{ptc5} \end{eqnarray}
The driving is then chosen to have the property \beq \int_0^T dt ~h_x(t)
= \frac{\pi}{2}, \label{ptc6} \eeq which gives $X = \prod_i
\sigma_i^x$ and $M=2$. This implies that $h_x \sim 1/T$ and it is
also assumed that $h_z,J \ll 1/T$. Then $D=-J\sum_{\langle ij
\rangle}\sigma_i^z \sigma_j^z + \cdots$ where $\cdots$ denote
higher-order corrections that preserve the Ising symmetry since
$[D,X]=0$. Starting with a short-ranged correlated state
$|\psi(0)\rangle$ which breaks the Ising symmetry and which also has
an initial energy density (with respect to the Hamiltonian $D$) that
corresponds to a temperature $T<T_c$ in two dimensions (where $T_c$
denotes the critical temperature for spontaneous breaking of the
Ising symmetry) ensures that $\langle \sigma_i^z (2nT)\rangle =
-\langle \sigma_i^z ((2n+1)T) \rangle \neq 0$. Here, we have implicitly
assumed that the system locally ``thermalizes'' with respect to the
Hamiltonian $D$ starting from the initial state $|\psi(0)\rangle$ on a
timescale $t_{th} \ll t_{*}$. Thus, the discrete time
translation symmetry of the system is spontaneously broken which
results in a prethermal Floquet time crystal that eventually melts
away for times $t \gg t_*$. As long as the discrete time translation
symmetry of the drive is unbroken, this prethermal phase is stable to
any small perturbations in Eq.~\eqref{ptc5}.

\section{Adiabatic-Impulse Approximation}
\label{sec:adimp}

In this section, we discuss the adiabatic-impulse method. It is one
of the few methods which can compute the Floquet Hamiltonian
accurately in the low-frequency drive regime. In this sense, it is
complementary to the FM expansion described in the previous section.
A somewhat detailed account of this method has been presented in
Ref.\ \onlinecite{rev4}. Here we will briefly discuss its salient
features, chart out the basic computations involved, and discuss its
recent application to integrable periodically driven systems.

To this end, we first consider a two-level Hamiltonian given by
\begin{eqnarray} H_2 &=& \epsilon(t) \sigma_z + \Delta_0 \sigma_1,
\label{adimpham2level} \end{eqnarray}
where $\sigma_{x,z}$ are Pauli matrices and $\Delta_0$ is a
constant. We will consider $\epsilon(t)=\epsilon_0 f(\omega_D t)$
to be an arbitrary periodic function of time, characterized by a
drive amplitude $\epsilon_0$ and a periodic time-dependent function
$f(\omega_D t)$, where $\omega_D= 2\pi /T$ is the drive frequency,
and $T$ is the time period. The method yields an accurate description
of the system for $\epsilon_0^2 +\Delta_0^2 \gg (\hbar \omega_D)^2$
and is thus suitable for capturing the low-frequency drive regime.

The central quantity that one aims to obtain using this technique is
the unitary evolution operator $U(t,0)$ which maps the initial state
to the final state at time $t$: $|\psi(t)\rangle = U(t,0) |\psi_{\rm
in}\rangle$. The adiabatic-impulse approximation allows a
semi-analytic computation of $U(t,0)$ for all $t$ and thus is
suitable for describing the micromotion of the system. This also
means that it provides us information about the phase bands, or
instantaneous eigenvalues of $U$, of the system \cite{topo5}. This
feature and the applicability to low-frequency dynamics
distinguishes this method from most other approximate analytical
techniques for computing $H_F$.

\begin{figure}
\rotatebox{0}{\includegraphics*[width= 0.98 \linewidth]{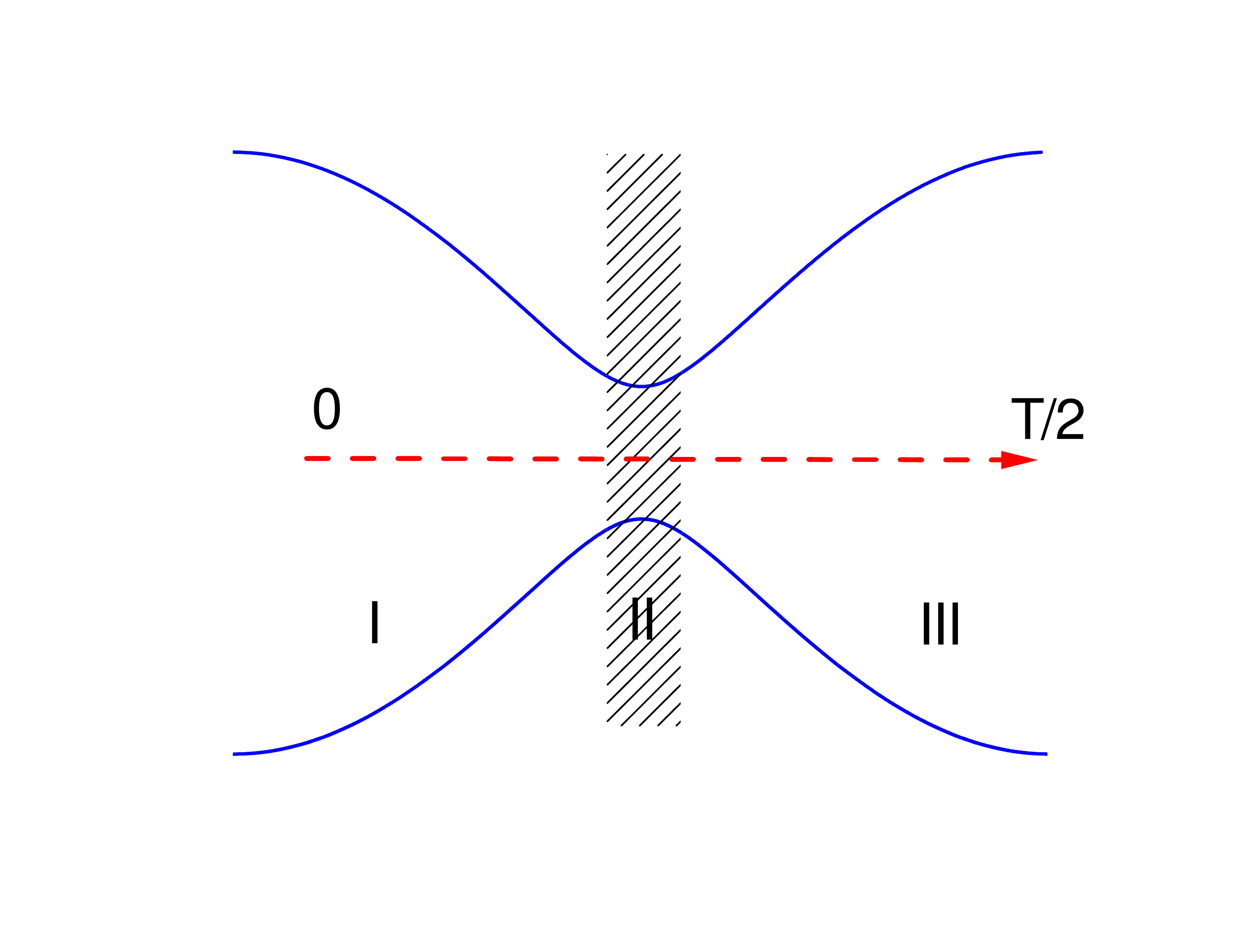}}
\caption{Schematic representation of the instantaneous energy levels
of a two-level system during its evolution from $t=0$ to $t=T/2$
with $\epsilon(t)= 5 \cos t$ and $\Delta=0.5$. Regions I and III
correspond to adiabatic evolution while region II (shaded area)
denotes the impulse region. The width of the impulse region is taken
to be close to zero in the adiabatic-impulse approximation. The red
line indicates the evolution of the system for the first half-cycle
from $t=0$ to $t=T/2$. The remaining half-cycle is traversed in the
opposite direction so that each region is traversed twice during
evolution from $t=0$ to $t=T$.} \label{figadimp1} \end{figure}

To chart out the computation of $U$ using this method, we first note
that the instantaneous eigenvalue of $U(t,0)$ can be trivially
found from Eq.\ \eqref{adimpham2level} and are given by
\begin{eqnarray} E_{\pm}(t) &=& \pm E(t), \quad E(t)= \sqrt{\epsilon(t)^2 +
\Delta_0^2}. \label{eigenval2level} \end{eqnarray}
The corresponding instantaneous eigenvectors are given by
\begin{eqnarray} |\psi_-(t)\rangle &=& (u(t),v(t))^T, \,\, |\psi_+(t)\rangle
= (-v(t),u(t))^T, \nonumber\\
u(t)&=& \frac{\Delta_0}{D(t)},\quad v(t)= (E(t)+\epsilon(t))/D(t), \nonumber\\
D(t) &=& \sqrt{[E(t)+\epsilon(t)]^2+\Delta_0^2}. \label{eigenfn2level}
\end{eqnarray}
A plot of this instantaneous energy gap $\delta E(t)= 2E(t)$ is
shown in Fig.\ \ref{figadimp1}. The plot clearly indicates that the
evolution time may be divided into two distinct regimes. In the
first regime, shown in Fig.\ \ref{figadimp1} as regions I and
III, one has $\delta E^2(t)/|\hbar\, d \delta E(t)/dt| \gg 1$; thus
as per standard Landau criterion, the system in these regions
undergo near-adiabatic evolution. In the other regime, denoted in
Fig.\ \ref{figadimp1} as region II, $\delta E^2(t)/|\hbar\, d \delta
E(t)/dt| \le 1$ and the evolution leads to the production of
excitations. This is the impulse region. The key approximation of the
adiabatic-impulse technique which enables one to analytically compute
$U$ is to treat the impulse region as one with an infinitesimal
width around a minimum of the instantaneous energy gap. Since this
approximation clearly becomes better at lower drive frequencies, the
adiabatic-impulse method naturally describes the low drive-frequency
regime accurately.

To compute $U(t,0)$ we note that the wave function of the system at
any time can be expressed in the adiabatic or the moving basis as
\begin{eqnarray}
|\psi(t)\rangle &=& c_1(t) \left( \begin{array}{c} u(t) \\ v(t)
\end{array} \right) + c_2(t) \left(\begin{array}{c} -v(t) \\ u(t)
\end{array}\right), \label{adbasiseq} \end{eqnarray}
where $c_1(0)=1$ and $c_2(0)=0$. The choice of this basis makes
the computation simpler, specially in the adiabatic regions I and III. We
note that the basis vectors
\begin{eqnarray} |\psi_{\rm ad}^{(0)}\rangle &=& (u(t),v(t))^T,\nonumber\\
|\psi_{\rm ad}^{(1)}\rangle &=& (-v(t),u(t))^T, \label{bvector} \end{eqnarray}
are related to those in the diabatic basis (given by $|\psi_{\rm
dia}^{(0)}\rangle = (u(t=0),v(t=0))^T$ and $|\psi_{\rm
dia}^{(1)}\rangle = (-v(t=0),u(t=0))^T$) by the standard transformation
\begin{eqnarray}
\left(\begin{array}{c} |\psi_{\rm ad}^{(0)}\rangle \\ |\psi_{\rm
ad}^{(1)}\rangle \end{array} \right) &=& \Lambda(t)
\left(\begin{array}{c}
|\psi_{\rm dia}^{(0)}\rangle\\
|\psi_{\rm dia}^{(1)}\rangle \end{array}
\right), \nonumber\\
\Lambda(t) &=& \left(\begin{array}{cc} \eta & \sqrt{1-\eta^2} \\
-\sqrt{1-\eta^2} & \eta \end{array} \right), \nonumber\\
\eta \equiv \eta(t) &=& u(t) u(0) + v(t) v(0). \label{addiatrans} \end{eqnarray}
We note that the adiabatic and the diabatic basis coincide at $t=0$
where $\eta=1$.

In region I, as discussed above, the system does not produce any
excitations. Thus the dynamics leads to a kinetic phase. This can be
seen most simply in the adiabatic basis where a straightforward
calculation, charted out in Refs.\ \onlinecite{rev4},
\onlinecite{topo4}, and \onlinecite{graphene1} shows
\begin{eqnarray}
c_1(t) &=& \exp[-i \zeta(t,0)] c_1(0), \nonumber \\
c_2(t) &=& \exp[i \zeta(t,0)] c_2(0), \nonumber\\
\zeta(t_1,t_2) &=& \frac{1}{\hbar} \int_{t_2}^{t_1} dt' E(t'). \label{cexp}
\end{eqnarray}
Thus in this basis one can define $U'_I(t,0) = \exp[-i \sigma_z
\zeta(t,0)]$ which relates $(c_1(t),c_2(t))^T$ to their values at $t=0$,
\begin{eqnarray}
\left(\begin{array}{c} c_1(t) \\ c_2(t)
\end{array} \right) &=& U'_{I}(t,0) \left(\begin{array}{c}
c_1(0)\\ c_2(0) \end{array} \right). \label{ceq1} \end{eqnarray}
Note that although $U'_I(t,0)$ is not the true evolution operator,
it acts as an useful operator which provide a handy calculational tool in
the adiabatic basis. To find the true evolution operator $U_I(t,0)$ for all
$t$ where the system is in region I, we use Eq.\ \eqref{addiatrans} to obtain
\begin{eqnarray} |\psi_I(t)\rangle &=& \Lambda(t) U'_I(t,0) |\psi_{\rm in}
\rangle = U_I(t,0) |\psi_{\rm in}\rangle, \label{wavreg1} \end{eqnarray}
where in the last line we have used the definition
$|\psi_I(t)\rangle = U_I (t,0) |\psi_{\rm in}\rangle$. This finally yields
\begin{eqnarray} U_I(t) &=& \left( \begin{array}{cc} e^{-i \zeta(t,0)}
\eta(t) &- e^{i \zeta(t,0)} \sqrt{1-\eta^2} \\
e^{-i \zeta(t,0)} \sqrt{1-\eta^2} & e^{i \zeta(t,0)} \eta(t)
\end{array} \right), \label{uevolreg1} \end{eqnarray}
which allows us to track the time evolution of the system in region
I. A similar calculation holds for any adiabatic region.

Next, we consider region II which is reached at $t=t_0$ as shown in
Fig.\ \ref{figadimp1}. Here the drive leads to the production of
defects. The width of this region, $\Delta t$, is approximated to be
infinitesimal within the adiabatic-impulse approximation. The width
of region II can be computed from the Landau criteria $\delta E^2
\le \hbar |d \delta E/dt|$; since $|d\delta E /dt| \sim \omega_D$,
it is clear that $\Delta t$ decreases with decreasing $\omega_D$.
Thus this approximation becomes better with decreasing $\omega_D$.
Typically one assumes that the width of this region is small enough
so that the wave functions immediately before entering region II and
immediately after leaving it are related by a transfer matrix
${\mathcal N}$
\begin{eqnarray} \left(\begin{array}{c} c_1(t_0+\Delta t) \\ c_2(t_0+\Delta t)
\end{array} \right) &=& {\mathcal N} \left(\begin{array}{c} c_1(t_0-\Delta t)\\
c_2(t_0-\Delta t) \end{array} \right). \label{ceq2} \end{eqnarray}
To compute ${\mathcal N}$, one typically uses a linearized
description of $H$. Within this approximation one writes $H(t)
\simeq [\epsilon(t_0) + (t-t_0) \dot \epsilon(t_0)]\sigma_z +
\sigma_x \Delta_0$, where $t_0$ is the time at which the system
reaches region II. The linearization of $H(t)$ around $t=t_0$
reduces the problem to that of computing the probability of the
generation of defects due to Kibble-Zureck mechanism \cite{rev7}. It
is well-known that the probability $p$ of defect formation in this
case is given by
\begin{eqnarray} p &=& \exp[-2 \pi \delta], \quad \delta= \Delta_0^2/[2 \hbar
\dot \epsilon_0(t_0)]. \label{kz1} \end{eqnarray} So for the
two-level system, the probability for the system to remain in its
starting state after crossing the impulse region is $1-p$. Thus the
diagonal elements of ${\mathcal N}$ yields $N_{11}, N_{22} \sim
\sqrt{1-p}$ while its off-diagonal element satisfies $N_{12},N_{21}
\sim \sqrt{p}$. The detailed computation of ${\mathcal N}$ from
these considerations has been charted out in Refs.\
\onlinecite{rev4}, \onlinecite{graphene1}, \onlinecite{child1} and
\onlinecite{kaya1} and yields
\begin{eqnarray} {\mathcal N} &=& \left(\begin{array} {cc} \sqrt{1-p}
e^{-i \phi_0} &
-\sqrt{p} \\ \sqrt{p} & \sqrt{1-p} e^{i \phi_0} \end{array} \right),
\nonumber\\
\phi_0 &=& \phi_{\rm st} -\pi, \label{transfer1} \\
\phi_{\rm st} &=& \frac{\pi}{4} + \delta \ln \delta + {\rm
Arg}\Gamma[1- i \delta], \nonumber \end{eqnarray}
where $\phi_{\rm st}$ is the Stoke's phase and $\Gamma$ denotes the
gamma function.

At the end of region II, one can write
\begin{eqnarray} \left(\begin{array}{c} c_1(t_0+\Delta t) \\ c_2(t_0+\Delta t)
\end{array} \right) &=& {\mathcal N} U'_I(t_0-\Delta t,0)
\left(\begin{array}{c} 1\\ 0\end{array} \right). \end{eqnarray}
Thus the evolution operator after the system has traversed region II
is given by
\begin{eqnarray} U_{II}(t_0+\Delta t,0) &=& \Lambda(t_0+\Delta t) {\mathcal N}
U'_I(t_0-\Delta t,0). \label{reg2} \end{eqnarray}
This procedure can be continued to obtain $U(t,0)$ for all $t \le
T$. To this end, we note that the system crosses the impulse region
twice, at $t=t_0$ and $t=T-t_0$; the rest of the dynamics consists of
passing through adiabatic regions. The evolution operator during
any time $t$ can be written as
\begin{eqnarray}
U(t,0) &=& \Lambda(t) \, U'_I(t,0), \quad t \le t_0 \label{evolop} \\
&=& \Lambda(t) \, U'_I(t,t_0) \, {\mathcal N} \, U'_I(t_0,0), \quad
T-t_0 \le t \le t_0 \nonumber\\
&=& \Lambda(t)\, U'_I(t, T-t_0)\, {\mathcal N}^T \, U'_I(T-t_0,t_0)
\nonumber\\
&& \times ~{\mathcal N} \, U'_I(t_0,0), \quad T \le t \le T_0-t.
\nonumber
\end{eqnarray}
where ${\mathcal N}^T$ denotes the transpose of ${\mathcal N}$. Thus
this method may be used to compute $U(t,0)$ for all $t \le T$ and
thus obtain information about the micromotion.

The instantaneous eigenvalues $\lambda_{\pm}(t)=\exp[\pm i
\theta(t)]$ of the evolution operator $U(t,0)$ are called phase
bands. They play an important role in charting out possible
topological transitions in driven many-body systems
\cite{topo1,topo5}. Moreover, at $t=T$, one can read off the
eigenvalues of the Floquet Hamiltonian from them: $\lambda_{\pm}(T)=
\exp[\pm i \epsilon_F T/\hbar]$. A straightforward computation,
charted out in Refs.\ \onlinecite{rev4} and \onlinecite{graphene1} yields
\begin{widetext}
\begin{eqnarray}
\cos \theta(t) &=& \eta \cos \alpha_1(t), \quad t \le t_0 \nonumber\\
&=& \eta \sqrt{1-p} \cos \alpha_2(t) + \sqrt{p(1-\eta^2)} \cos
\beta_2(t), \quad t_0 \le t \le T-t_0 \label{pband1} \\
&=& \eta[p \cos \alpha_3(t) +(1-p) \cos \beta_3(t)] +
\sqrt{p(1-p)(1-\eta^2)} [\cos (\alpha_3 - \phi_{\rm st}) -
\cos(\beta_3 - \phi_{\rm st}) ], \quad t \ge T_0-t \nonumber
\end{eqnarray}
\end{widetext}
where we have diagonalized $U(t,0)$ in Eq.\ \eqref{evolop} to obtain
these expressions, and $\alpha_{1,2,3}$ and $\beta_{2,3}$ are given by
\begin{eqnarray}
\alpha_1(t) &=& \zeta(t,0), \quad \alpha_2(t)= \zeta(t,0)+\phi_{\rm
st}, \nonumber\\
\beta_2(t) &=& \zeta(t,t_0)-\zeta(t_0,0), \quad \alpha_3(t)= 2
\phi_{\rm st} + \zeta(t,0), \nonumber\\
\beta_3(t) &=& \zeta(t_0,0)-\zeta(T-t_0,t_0)+\zeta(t,T-t_0).
\label{phasefac} \end{eqnarray}

The computational scheme charted above brings out two aspects of the
method. First, it can be directly applied to a class of integrable
spin models which can be written in terms of free fermions via a
Jordan-Wigner transformation. These models include the
one-dimensional Ising and XY models and the two-dimensional Kitaev models
\cite{isingref, kitaevref}. In addition, it can also be used to
describe the dynamics of Dirac quasiparticles in graphene or atop a
topological insulator surface \cite{grapheneref, tiref}, and Weyl
fermions in 3D band systems \cite{weylref}. All these systems can
be represented by fermionic Hamiltonians of the form
\begin{eqnarray}
H &=& \sum_{\vec k} \psi_{\vec k}^{\dagger} H_{\vec k} \psi_{\vec k},
\label{hamdir1} \end{eqnarray}
where $H_{\vec k}$ is given by Eq.\ \eqref{adimpham2level} with
$\epsilon_0(t) \to \epsilon_{\vec k}(t)$ and $\Delta_0 \to
\Delta_{\vec k}$. The precise forms of $\epsilon_{\vec k}(t)$ and
$\Delta_{\vec k}$ depend on the model and are well-known
\cite{isingref, kitaevref,grapheneref,tiref,weylref}. Second, the method
provides an easy access to the micromotion in these systems; thus it allows
one to address the phase bands of these models. It has been recently pointed
out that the understanding of topological transitions in such driven systems
requires an analysis of their phase bands $\theta(\vec k, t)$, and a
knowledge of only their Floquet spectrum $\epsilon_{F}(\vec k) =
\theta(\vec k, T)/T$ may not be sufficient~\cite{topo5}.
We note in passing that this scheme can be generalized to cases
where both $\epsilon$ and $\Delta$ are time dependent; the details
of such generalizations have been charted out in Refs.\
\onlinecite{graphene1} and \onlinecite{sau1}.

In what follows, we will provide an example of graphene in the
presence of external radiation where one can use this method to
detect a topological transition at $t=T/3$ \cite{graphene1}. The
Hamiltonian of graphene in the presence of an external radiation is
given by Eq. \eqref{hamdir1} with $\epsilon_{\vec k}(t) = - {\rm Re}
Z_{\vec k}(t)$ and $\Delta_{\vec k}(t)= {\rm Im} Z_{\vec k}(t)$, where
\begin{eqnarray}
Z_{\vec k}(t) &=& -\sum_{p=\pm 1} e^{ i (\cos(\omega t- p \pi/3) +
(k_x + \sqrt{3} p k_y)/2)} \nonumber\\
&& - ~e^{ i(\cos(\omega t) -k_x)}, \label{zfn} \end{eqnarray}
where $\alpha=e A_0/c$, and $A_0$ and $\omega$ are the amplitude and
frequency of the circularly polarized external radiation represented
by the vector potential $\vec A= A_0 (\cos(\omega t), \sin(\omega
t))$. It can be directly checked that at the $\Gamma$ point of the
Brillouin zone ($(k_x,k_y)=(0,0)$), $H_{\vec k}$ satisfies
\begin{eqnarray} H_{\vec k}(t) &=& H_{\vec k}(T-t), \label{symmcond} \\
H_{\vec k}(T/3 \pm t) &=& H_{\vec k}(t) = H_{\vec k} (2T/3 \pm t),
\nonumber\\
U(2T/3,0) &=& [U(T/3,0)]^2, \quad U(T,0)= [U(T/3,0)]^3, \nonumber
\end{eqnarray}
where $U$ represents the evolution operator at the $\Gamma$ point.
This shows that a phase band crossing leading to a change of
topology of the driven system at $t=T/3$ (which amounts to
$U(T/3,0)=\pm I$) necessarily shows analogous crossing at $t=T$;
however, the reverse is not true.

The verification of such crossings at $t=T/3$ and $2T/3$ has been carried
out in detail in Ref.\ \onlinecite{graphene1}. A somewhat lengthy calculation
yields an analytical expression for the phase bands within the
adiabatic-impulse approximation. In terms of the probability $p_{\Gamma}$
for the formation of excitations formation probability and the associated
Stuckelberg phase $\Phi_{\Gamma}$, one finds that the expression for
the phase band $\phi_{\Gamma}$ at the $\Gamma$ point is \cite{graphene1}
\begin{eqnarray} \cos(\phi(T/3)) &=& p_{\Gamma} + (1-p_{\Gamma}) \cos(2
\Lambda_{\Gamma}), \label{pcond1} \\
\Lambda_{\Gamma} &=& \Phi_{\Gamma} + 2 \int_0^{T/6} dt
\sqrt{\epsilon_{\vec k=0}^2(t) + \Delta_{\vec k=0}^2(t)}. \nonumber
\end{eqnarray}
It was shown that the band crossings that lead to a change in
topology of the state of the driven system at $t=T/3$ requires
$\cos[\phi_{\Gamma}(T/3)]=+(-)1$ for crossings through the zone
center (edge). The crossings through the zone center thus requires
$\Lambda_{\Gamma} = m\pi$ for $m \in Z$. The crossing through the
zone edge, in contrast, necessitates $ \cos[2 \Lambda_{\Gamma}] = -(
1+p_{\Gamma})/(1-p_{\Gamma})$; this is clearly untenable for real
$\Lambda$ and hence the adiabatic-impulse approximation predicts
that all such band crossings at $t=T/3$ should occur through the
zone center. This fact has been numerically verified in Ref.\
\onlinecite{graphene1}. A similar analysis has been carried out for other
values of $T$ and at other points in the graphene Brillouin zone. In
all cases, the prediction of the adiabatic-impulse method provides a
near-exact match with exact numerics as long as the drive
frequency is small compared to the nearest-neighbor hopping
amplitude of the electrons in graphene; in addition it provides
analytical conditions for phase band crossings which help in
obtaining a semi-analytic understanding of the phase diagram of periodically
driven graphene \cite{graphene1}. Moreover, such an analysis can be easily
extended to a wide class of driven spin and fermionic systems which
host Dirac-like quasiparticles. It thus provides a complete picture
of the low-frequency behavior of a wide range of integrable models.

\section{Floquet perturbation theory}
\label{sec:fpt}

In this section, we discuss a perturbative method to find the
Floquet Hamiltonian $H_F$ or periodically driven many-body
Hamiltonians of the form $H(t)=H_0(t)+ g V(t)$, where $H_0(t) (V(t))
= H_0(t+T) (V(t+T))$ (note that $H_0(t)$ or $V(t)$ may be
time-independent), $g \ll 1$, and crucially, $H_0(t)$ consists of
mutually commuting terms. We call this Floquet perturbation theory
(FPT) whereby $H_F$ is obtained as a power series in
$g$~\cite{soori10,flscar1,haldar19}. This method is particularly
suited to address the nature of the Floquet Hamiltonian at
intermediate and low drive frequencies, unlike the high-frequency FM
expansion.

As the first example, suppose that the Hamiltonian $H$ can be written as a
sum of two parts, $H(t)$ which varies periodically in time with a period
$T = 2 \pi/\omega_D$, and a perturbation $V$ which is time-independent.
Thus $H(t) = H_0 (t) + V$. Since $H_0 (t)$ commutes with
itself at different times, we can work in the basis of
eigenstates of $H_0 (t)$ which are time-independent and orthonormal.
We denote these as $| n \rangle$, so that $H_0 (t) | n \rangle = E_n
(t) | n \rangle$, and $\langle m | n \rangle = \delta_{mn}$.

We now find solutions of the time-dependent Schr\"odinger equation
\beq i \hbar \frac{\partial \psi_n}{\partial t} ~=~ H(t) \psi_n (t)
\label{sch1} \eeq which satisfy the Floquet eigenstate condition
\beq \psi_n (T) ~=~ e^{i \theta_n} ~\psi_n (0), \label{floeig1} \eeq
where $e^{i \theta_n}$ is the Floquet eigenvalue.

For $V=0$, we have \beq \psi_n (t) ~=~ e^{-(i/\hbar) \int_0^t dt'
E_n (t')} ~|n \rangle, \eeq so that the eigenvalue of the Floquet unitary
\beq U ~=~ {\cal T} ~e^{-(i/\hbar) \int_0^T dt H(t)} \label{flo1} \eeq
is given by \beq e^{i \theta_n} ~=~ e^{-(i/\hbar) \int_0^T dt E_n (t)}.
\label{floeig2} \eeq For $V$ non-zero but small, we will develop a
FPT to first order in $V$. We first consider
non-degenerate perturbation theory; the meaning of non-degenerate
will become clear below. We assume that the $n$-th eigenstate can be
written as \beq \psi_n (t) ~=~ \sum_m ~c_m (t) ~e^{-(i/\hbar)
\int_0^t dt' E_m (t')} ~|m \rangle, \label{psi1} \eeq where $c_n (t)
= 1 ~+$ terms of order $V$ for all $t$, while $c_m (t)$ is of order
$V$ for all $m \ne n$ and all $t$. Eq.~\eqref{sch1} then implies
\bea && i \hbar ~\sum_m \dot{c}_m (t) e^{-(i/\hbar) \int_0^t dt' E_m
(t')} ~|m \rangle \nonumber \\
&=& V ~\sum_m ~c_m (t) ~e^{-(i/\hbar) \int_0^t dt' E_m (t')} ~|m
\rangle, \label{sch2} \eea where $\dot{c}_m$ denotes $dc_m/dt$.
Taking the inner product of Eq.~\eqref{sch2} with $\langle n |$, we
find, to first order in $V$, that $i \hbar \dot{c}_n = \langle n |V
| n \rangle$. Choosing $c_n (0) = 1$, we then have \beq c_n (t) ~=~
e^{-(i/\hbar) \langle n |V | n \rangle t}. \label{cnt1} \eeq This
gives \bea \psi_n (t) &=& e^{-(i/\hbar) (\langle n |V | n \rangle t
~+~ \int_0^t dt' E_n (t'))} ~|n \rangle \nonumber \\
&& +~ \sum_{m \ne n} ~c_m (t) ~e^{-(i/\hbar) \int_0^t dt' E_m (t')}
~|m \rangle. \label{psi2} \eea

Next, taking the inner product of Eq.~\eqref{sch2} with $\langle m
|$, where $m \ne n$, we find, to first order in $V$, that \beq
\dot{c}_m ~=~ - ~\frac{i}{\hbar} ~\langle m | V | n \rangle
~e^{(i/\hbar) \int_0^t dt' [E_m (t') - E_n (t')]}. \label{cmdot}
\eeq (We have ignored a factor of $e^{(i/\hbar) \langle n |V | n
\rangle t}$ on the right hand side of Eq.~\eqref{cmdot} since we are
only interested in terms of first order in $V$). Integrating
Eq.~\eqref{cmdot} gives \beq c_m (T) ~=~ c_m (0) ~-~ \frac{i}{\hbar}
~\langle m | V | n \rangle ~ \int_0^T dt ~e^{i \int_0^t dt' [E_m
(t') - E_n (t')]}. \label{cmt1} \eeq Since we know that
Eq.~\eqref{psi2} satisfies \beq \psi_n (T) ~=~ e^{-(i/\hbar)
(\langle m | V | n \rangle T ~+~ \int_0^T dt E_n (t))} ~\psi_n (0),
\eeq we must have, to first order in $V$, \beq c_m (T) ~=~
e^{(i/\hbar) \int_0^T dt [E_m (t) - E_n (t)]} ~c_m (0) \eeq for all
$m \ne n$. This, along with Eq.~\eqref{cmt1}, means that we must
choose \beq c_m (0) ~=~ - \frac{i}{\hbar} ~\langle m | V | n \rangle
~\frac{\int_0^T dt ~e^{(i/\hbar) \int_0^t dt' [E_m (t') - E_n
(t')]}}{e^{(i/\hbar) \int_0^T dt [E_m (t) - E_n (t)]} ~-~ 1}.
\label{cmt2} \eeq We see that $c_m (t)$ is indeed of order $V$
provided that the denominator on the right hand side of
Eq.~\eqref{cmt2} does not vanish; we call this case non-degenerate.
If \beq e^{(i/\hbar) \int_0^T dt [E_m (t) - E_n (t)]} ~=~ 1,
\label{res1} \eeq we have a resonance between states $| m \rangle$
and $|n \rangle$, and the above analysis breaks down. We then have
to develop a degenerate FPT as discussed below.

If there are several states which are connected to $|n\rangle$ by
the perturbation $V$, Eq.~\eqref{cmt2} describes the amplitude to go
to each of them from $|n\rangle$. Up to order $V^2$, the total
probability of excitation away from $|n\rangle$ is given by $\sum_{m
\ne n} |c_m (0)|^2$. If $c_m (0)$ turns out to be zero for all $m
\ne n$ (this can happen if either the matrix element $\langle m | V
| n \rangle = 0$ or the numerator of the expression in
Eq.~\eqref{cmt2} vanishes), the Floquet eigenstate remains equal to
$\psi_n$ up to first order in $V$. This is an example of dynamical freezing.

Next we consider degenerate perturbation theory. Suppose that there
are $p$ states $|m\rangle$ ($m=1,2,\cdots,p$) which have energies
$E_m$ and satisfy Eq.~\eqref{res1} for every pair of states $(m,n)$
lying in the range 1 to $p$. Ignoring all the other states for the
moment, we assume that a solution of the Schr\"odinger equation is
given by \beq \psi_n (t) ~=~ \sum_{m=1}^p ~c_m (t) ~e^{-(i/\hbar)
\int_0^t dt' E_m (t')} ~ |m \rangle, \label{psit} \eeq where we now
allow all the $c_m (t)$'s to be of order 1. To first order in $V$, we
can then replace $c_m (t)$ by the time-independent constants $c_m
(0)$ on the right hand side of Eq.~\eqref{sch2}. Upon integrating
from $t=0$ to $T$, we obtain \bea c_m (T) &=& c_m (0) ~-~
\frac{i}{\hbar} ~\sum_{n=1}^p ~\langle m | V | n \rangle \nonumber \\
&& \times \int_0^T dt ~e^{(i/\hbar) \int_0^t dt' [E_m(t') -
E_n(t')]} ~c_n (0). \label{cmt3} \eea This can be written as a
matrix equation \beq c(T) ~=~ [I ~-~ \frac{iH' T}{\hbar}] ~c(0),
\label{ct1} \eeq where $c(t)$ denotes the column
$(c_1(t),c_2(t),\cdots,c_p(t))^T$ (where the superscript $T$ denotes
transpose), $I$ is the $p$-dimensional identity matrix, and $H'$ is
a $p$-dimensional Hermitian matrix with matrix elements \beq
(H')_{mn} ~=~ \frac{\langle m | V | n \rangle}{T} ~\int_0^T dt
e^{(i/\hbar) \int_0^t dt' [E_m(t') - E_n(t')]}. \label{mmn} \eeq Let
the eigenvalues of $H'$ be $\epsilon_n$ ($n=1,2,\cdots,p$). To first
order in $V$, $I - i H' T/\hbar$ is a unitary matrix and therefore
has eigenvalues of the form $e^{-i\epsilon_n T/\hbar}$; the
corresponding eigenstates satisfy \beq c (T) ~=~ e^{-i\epsilon_n
T/\hbar} ~c (0). \eeq Next, we want the wave function in
Eq.~\eqref{psit} to satisfy Eq.~\eqref{floeig1}. This implies that
the Floquet eigenvalues are related to the eigenvalues of $H'$ as
\beq e^{i\theta_n} ~=~ e^{-(i/\hbar) (\epsilon_n T ~+~ \int_0^T dt
E_n (t))}, \eeq where we have used the resonance condition that
$e^{-(i/ \hbar) \int_0^T dt E_n (t)}$ has the same value for all
$n=1,2,\cdots,p$.

Given a Floquet unitary $U(T)$, we can define a Floquet Hamiltonian
$H_F$ using Eq.~\eqref{HFdef}. 
Comparing this with
Eqs.~\eqref{ct1} and \eqref{mmn}, we see that the matrix elements of
$H_F$ are \bea (H_F)_{mn} &=& \frac{\langle m | V | n \rangle}{T} \int_0^T dt
e^{(i/\hbar) \int_0^t dt' [E_m(t') - E_n(t')]} \nonumber \\
&& + ~\Bigl( \frac{1}{T} \int_0^T dt ~E_n (t) \Bigr) ~\delta_{mn},
\label{hmn1} \eea where we have assumed that $\int_0^T dt ~E_n (t)$
has the same value for all $n$. [This is a special case of
Eq.~\eqref{res1}. More generally, Eq.~\eqref{res1} allows the values
of $\int_0^T dt ~E_n (t)$ to differ from each other by non-zero
integer multiples of $2 \pi \hbar$, but we will not consider that
possibility here].

Unlike the FM expansion~\cite{rev1,mikami16}, FPT does not assume
the drive frequency $\omega_D$ to be large compared to the other
parameters of the system. It only assumes the amplitude of the
driving to be large. This will become clear in the examples
discussed below where we will see that the Floquet Hamiltonian is
effectively an expansion in the inverse of the driving amplitude.

As the first application of the above formalism, we consider a
simple model with a single spin-1/2 which is governed by a
Hamiltonian $H(t) = H_0 (t) + V$, where~\cite{udupa20}
\bea H_0 &=& \lambda \cos (\omega t) ~\sigma^x, \nonumber \\
V &=& g_1 ~\sigma^x ~+~ g_2 ~\sigma^y, \label{hv} \eea and we will
assume that $\lambda \gg g_1, ~g_2$. The unperturbed problem given
by $i \hbar \partial \psi /\partial t = H_0 \psi$ has solutions \bea
\psi_1 (t) &=& e^{-i (\lambda / \hbar \omega) \sin (\omega t)}~
| 1 \rangle, \nonumber \\
\psi_2 (t) &=& e^{i (\lambda / \hbar \omega) \sin (\omega t)} ~| 2
\rangle,
\nonumber \\
{\rm where} ~~| 1 \rangle &=& \frac{1}{\sqrt{2}} \begin{pmatrix} 1
\\ 1 \end{pmatrix} ~~{\rm and}~~ | 2 \rangle ~=~ \frac{1}{\sqrt{2}}
\begin{pmatrix} 1 \\ -1 \end{pmatrix}, \eea
and the corresponding eigenvalues of $H_0 (t)$ are $E_1(t)= \lambda
\cos (\omega t)$ and $E_2(t)= - \lambda \cos (\omega t)$
respectively. Since these satisfy Eq.~\eqref{res1} we have to use
degenerate perturbation theory. Following
Eqs.~(\ref{psit}-\ref{hmn1}), and using the
identity~\cite{abramowitz} \beq \int_0^T dt ~e^{(i/\hbar) \int_0^t
dt' 2 \lambda \cos (\omega t')} ~=~ T J_0\Big({\frac{2\lambda}{\hbar
\omega}}\Big), \eeq we find that the Floquet Hamiltonian is \beq H_F
~=~ g_1 ~\sigma^x ~+~ g_2 ~J_0\Big({\frac{2\lambda}{\hbar
\omega}}\Big) ~\sigma^y. \label{ham3} \eeq The Bessel function $J_0
(z) \to 1$ when $z \to 0$ and goes as $\sqrt{2/(\pi z)} \cos (z -
\pi/4)$ when $z \to \infty$. It is clear that Eqs.~\eqref{hv} and
\eqref{ham3} are consistent with each other in the limit $\lambda
/(\hbar \omega) \to 0$; in particular, $H_F$ approaches the
time-averaged value $(1/T) \int_0^T dt H (t)$ in the high-frequency
limit. The limit $\lambda /(\hbar \omega) \to \infty$ is less
trivial; we then see that the driving-dependent term in $H_F$ goes
to zero as $\sqrt{\hbar \omega / \lambda}$ apart from an oscillatory
factor. Incidentally, we note that if we shift the time, i.e.,
change $\cos (\omega t) \to \cos (\omega (t+t_0))$ in
Eq.~\eqref{hv}, the expression for the Floquet Hamiltonian in
Eq.~\eqref{ham3} would change.

Note that if we had considered a different limit where $\hbar \omega
\gg \lambda, g$, and used the FM expansion, we would have
obtained an expansion in powers of $\lambda /(\hbar \omega)$ and $g
/(\hbar \omega)$. A resummation of all the terms which are of first
order in $g$ would then give back the expression in
Eq.~\eqref{ham3}. Thus first-order FPT gives an expression for $H_F$
which is a resummation of all the terms in the FM expansion
which are of first order in the perturbation $V$.

Next, we apply FPT to a periodically driven spin chain called the
PXP model~\cite{flscar1}. We consider $H(t) = H_0 (t) + V$, where
\bea H_0 (t) &=& \frac{\lambda (t)}{2} ~\sum_l ~\sigma^z_l, \nonumber \\
V &=& g ~\sum_l ~P_{l-1} \sigma_l^x P_{l+1}, \label{ham2} \eea where
$P_l = (1 - \sigma_l^z)/2$ is the projection operator to the
spin-down state at site $l$. The presence of the projection operator
in the second line of Eq.~\eqref{ham2} makes the Hamiltonian $V$
non-integrable. We will consider a driving with the form of a square pulse,
\bea \lambda (t) &=& - ~\lambda ~~~{\rm for}~~~ 0 ~<~ t ~<~ T/2 \nonumber \\
&=& +~ \lambda ~~~{\rm for}~~~ T/2 ~<~ t ~<~ T, \label{pulse} \eea
and $\lambda (t + T) = \lambda (t)$ for all times.

We now apply FPT assuming that $\lambda \gg g$. We choose the
$\sigma_l^z$ basis for the states. According to the unperturbed
Hamiltonian $H_0$ in Eq.~\eqref{ham2}, we see that such states $| n
\rangle$ have an instantaneous energy eigenvalue $E_n (t) = (\lambda
(t)/2) ~\sum_l ~\sigma^z_l$. We now consider the effect to first
order of the perturbation $V$ in Eq.~\eqref{ham2}. If $| m \rangle$
and $| n \rangle$ are two states which are connected by $V$, they
differ by the value of $\sigma_l^z$ at only one site and therefore
$E_m (t) - E_n (t) = \lambda (t)$. Hence Eq.~\eqref{res1} is
satisfied and we have to use degenerate perturbation theory. The
integral in Eq.~\eqref{cmt3} is found to be \beq \int_0^T dt
~e^{(i/\hbar) \int_0^t dt' [E_m(t') - E_n(t')]} ~=~ \frac{i2
\hbar}{\lambda} ~(e^{-i\lambda T/(2 \hbar)} ~-~ 1). \label{cond1}
\eeq We see that if $e^{-i \lambda T/(2 \hbar)} = 1$, i.e., if \beq
\frac{\lambda}{\hbar \omega} ~=~ 2p, \label{cond2} \eeq where $p$ is
an integer, then the expression in Eq.~\eqref{cond1} vanishes. This
means that even in degenerate perturbation theory, there is no
change in the Floquet eigenvalues and they remain equal to 1.

We can now use Eqs.~\eqref{hmn1} and \eqref{cond1} to derive the
Floquet Hamiltonian. If $|m \rangle$ and $| n \rangle$ are two
states which are connected by the perturbation $V$, we have $\langle
m | V | n \rangle = g$ (note that $|m \rangle$ and $| n \rangle$
must necessarily be different from each other). We then obtain \bea
(H_F)_{mn} &=& \frac{i2g \hbar}{\lambda T} ~(e^{-i\lambda T/(2\hbar)}
~-~ 1) \nonumber \\
&=& \frac{2g \hbar \omega}{\pi \lambda} ~e^{-i \pi \lambda/(2 \hbar
\omega)} ~\sin \Bigl( \frac{\pi \lambda}{2 \hbar \omega} \Bigl).
\eea We thus see that \beq H_F ~=~ g ~\frac{\sin \gamma}{\gamma}
~\sum_l ~P_{l-1} [\cos \gamma ~ \sigma_l^x ~+~ \sin \gamma
~\sigma_l^y] P_{l+1}, \label{ham4} \eeq where $\gamma = \pi \lambda
/(2 \hbar \omega)$. We now see that if Eq.~\eqref{cond2} holds, the
Floquet Hamiltonian vanishes to first order in $g/\lambda$. We then
have to go to higher orders or study the model numerically to
understand its behavior~\cite{flscar1}.

Eq.~\eqref{ham4} shows that in the limit $\lambda \to \infty$, $H_F$
goes to zero as $1/\lambda$ apart from some oscillatory factors. The
different power laws, $1/\sqrt{\lambda}$ versus $1/\lambda$, in
Eqs.~\eqref{ham3} and \eqref{ham4} are related to the fact that the
driving term has different forms in the two cases, $\lambda \cos
(\omega t)$ in the first case and the square pulse in
Eq.~\eqref{pulse} in the second case.

Finally, we consider a case where the Floquet Hamiltonian $H_F$ has
no contributions to first order in the perturbation and we have to
go up to second order. Further, we will take the periodically driven
part of the Hamiltonian to be much smaller than the time-independent
part, and we will calculate $H_F$ only within a particular sector of
eigenstates of the time-independent part. We consider a system with
$M$ states $|\alpha \rangle$ in sector 1, all of which have energy
$E_1$, and $N$ states $|\beta \rangle$ in sector 2, all of which
have energy $E_2$. We introduce a small time-dependent coupling
between states in the two sectors given by $V(t)$ and its Hermitian
conjugate $V^\dagger (t)$ as follows, where $V(t+T)=V(t)$. Denoting
states in sectors 1 and 2 by $\Psi_1$ and $\Psi_2$, which are
columns with $M$ and $N$ entries respectively, the Schr\"odinger
equation takes the form \bea i \hbar \frac{\partial \Psi_1}{\partial
t} &=& E_1 ~\Psi_1 ~+~ V ~\Psi_2, \nonumber \\
i \hbar \frac{\partial \Psi_2}{\partial t} &=& E_2 ~\Psi_2 ~+~
V^\dagger ~ \Psi_1, \label{sch4} \eea where $V$ is a $M \times N$
dimensional matrix. We now look for a Floquet eigenstate $\psi (t)$
which lies mainly in sector 1, namely, \bea \psi (t) &=& \left(
\sum_\alpha c_\alpha (t) |\alpha \rangle \right)
e^{-i E_1 t /\hbar} \nonumber \\
&& +~ \left( \sum_\beta c_\beta (t) |\beta \rangle \right) e^{-i E_2
t /\hbar}, \label{psi4} \eea where $c_\alpha (t)$ are of order 1
while $c_\beta (t)$ are of order $V$. Within sector 1, $H_F$ will be
equal to $E_1$ plus terms of order $V^2$ (there are no contributions
to first order in $V$ since the driving term has no matrix elements
within sector 1). To calculate $H_F$, we proceed as follows.
Denoting the column of coefficients $c_\alpha$ and $c_\beta$
collectively as $c_1$ and $c_2$ respectively, we have
\bea i \hbar {\dot c}_1 &=& e^{i(E_1 - E_2) t/\hbar} ~V ~c_2, \nonumber \\
i \hbar {\dot c}_2 &=& e^{i (E_2 - E_1) t/\hbar} V^\dagger ~c_1.
\label{sch5} \eea In the second equation in Eq.~\eqref{sch5}, we set
$c_1 (t) = c_1 (0)$ on the right hand side since we want to find
$c_2 (t)$ only to order $V$. Integrating in time, we obtain \beq c_2
(t) ~-~ c_2 (0) ~=~ - \frac{i}{\hbar} ~\int_0^t dt' e^{i (E_2 -
E_1)t' /\hbar} V^\dagger (t') c_1 (0). \label{ct2} \eeq Since the
Floquet eigenvalue in sector 1 is $e^{-iE_1 t/\hbar}$ to first order
in $V$, we require \beq c_1 (T) ~=~ c_1 (0) ~~~{\rm and}~~~ c_2 (T)
~=~ e^{i (E_2 - E_1) t/\hbar} c_2 (0) \label{bc1} \eeq to first
order. Using this, we find from Eq.~\eqref{ct2} that \bea c_2 (t)
&=& - \frac{i}{\hbar} ~[\frac{\int_0^T dt' e^{i (E_2 - E_1)
t'/\hbar} V^\dagger (t')}{e^{i (E_2 - E_1) T/\hbar} -1} \nonumber \\
&& ~~~~~~+ \int_0^t dt' e^{i (E_2 - E_1) t'/\hbar} V^\dagger (t')]
c_1 (0). \label{ct3} \eea The first equation in Eq.~\eqref{sch5},
then gives \beq c_1 (T) ~=~ c_1 (0) ~-~ \frac{i}{\hbar} \int_0^T dt
~e^{i (E_1 - E_2)t /\hbar} V(t) c_2 (t). \label{ct4} \eeq Since the
Floquet Hamiltonian in sector 1 satisfies $\psi_1 (T) = e^{-i H_F T/
\hbar} \psi_1 (0)$, we see from Eq.~\eqref{ct4} that \bea H_F &=&
E_1 I ~-~ \frac{i}{\hbar T} \int_0^T dt ~e^{i (E_1 - E_2)t /\hbar}
~V(t) \nonumber \\
&& \times ~[ \frac{\int_0^T dt' e^{i (E_2 - E_1)t' /\hbar}
~V^\dagger (t')}{
e^{i (E_2 - E_1) T/\hbar} -1} \nonumber \\
&& ~~~~+ ~\int_0^t dt' ~^{i (E_2 - E_1)t' /\hbar} ~V^\dagger (t') ].
\label{hf1} \eea Next, the time periodicity of $V$ allows us to
write it as \beq V ~=~ \sum_{n=-\infty}^\infty ~V_n e^{-in \omega
t}. \label{vm} \eeq Eq.~\eqref{hf1} then gives \beq H_F ~=~ E_1 I
~-~ \sum_{n=-\infty}^\infty \frac{V_n V_n^\dagger}{E_2 - E_1 + n
\hbar \omega}. \label{hf2} \eeq Eq.~\eqref{hf2} shows that
resonances occur whenever $(E_2 - E_1)/(\hbar \omega)$ is equal to
an integer. However, near these points the above derivation of $H_F$
breaks down since the Floquet eigenstates will no longer have $c_2
(t)$ much smaller than $c_1 (t)$, which was an assumption made in
order to calculate $H_F$. We note in passing that under a time shift
$t \to t + t_0$, we would have $V_n \to V_n e^{-in \omega t_0}$ in
Eq.~\eqref{vm}, but the Floquet Hamiltonian in Eq.~\eqref{hf2} would
{\it not} change.

As an example of the above formalism, we consider the Hubbard model
with two sites, labeled 1 and 2, where the hopping amplitude between
the two sites, $g$, is much smaller than the on-site interaction
strength $U$.~\cite{itin15} We will take the phase of the hopping to
be a sinusoidal function of time; this describes the effect of a
periodically varying electric field through the Peierls prescription
(see Sec.~\ref{ssec:rwa}). The Hamiltonian is \bea H &=& U
~\sum_{n=1,2} ~c_{n\uparrow}^\dagger c_{n\uparrow}
c_{n\downarrow}^\dagger c_{n\downarrow} \nonumber \\
&& - ~g~ \sum_{\sigma=\uparrow,\downarrow} ~(e^{\frac{ia}{\omega}
\sin (\omega t)} c_{1\sigma}^\dagger c_{2\sigma} \nonumber \\
&& ~~~~~~~~~~~~~~~+~ e^{-\frac{ia}{\omega} \sin (\omega t)}
c_{2\sigma}^\dagger c_{1\sigma}). \label{ham5} \eea We will consider
a half-filled system with two electrons. In the undriven system
($a=0$), we know that the low-energy states are described by an
effective spin Hamiltonian given by $(4 g^2/U) ({\vec S}_1 \cdot
{\vec S}_2 /\hbar^2 - 1/4)$. We will study what effect the driving
has on the effective Hamiltonian which will now be denoted by $H_F$.

In the space of two-electron states, the states \beq
|1\uparrow,2\uparrow \rangle ~=~ c_{1\uparrow}^\dagger
c_{2\uparrow}^\dagger | 0 \rangle ~~~{\rm and}~~~ |1\downarrow,
2\downarrow \rangle = c_{1\downarrow}^\dagger
c_{2\downarrow}^\dagger | 0 \rangle \label{2states} \eeq have a
trivial dynamics since $H$ annihilates these states. Hence they are
both Floquet eigenstates with Floquet eigenvalue equal to 1. Next,
we study the states in which there is one spin-up electron and one
spin-down electron. There are four such states, \bea |1\rangle &=&
|1 \uparrow, 2 \downarrow \rangle = c_{1\uparrow}^\dagger
c_{2\downarrow}^\dagger | 0 \rangle, \nonumber \\
|2 \rangle &=& |1 \downarrow, 2 \uparrow \rangle =
c_{1\downarrow}^\dagger
c_{2\uparrow}^\dagger | 0 \rangle, \nonumber \\
|3 \rangle &=& |1 \uparrow, 1 \downarrow \rangle =
c_{1\uparrow}^\dagger
c_{1\downarrow}^\dagger | 0 \rangle, \nonumber \\
|4 \rangle &=& |2 \uparrow, 2 \downarrow \rangle =
c_{2\uparrow}^\dagger c_{2\downarrow}^\dagger | 0 \rangle.
\label{4states} \eea In terms of Eqs.~\eqref{sch4}, the first two
states in Eq.~\eqref{4states} form sector 1 and have eigenvalues
$E_1 = 0$ (low energy), the last two states form sector 2 and have
eigenvalues $E_2 =U$ (high energy), and the matrix $V$ relating the
states of sector 2 to sector 1 is given by \bea V ~=~ g ~\left(
\begin{array}{cc} - e^{-\frac{ia}{\omega} \sin (\omega t)} & -
e^{\frac{ia}{\omega} \sin
(\omega t)} \\
e^{- \frac{ia}{\omega} \sin (\omega t)} & e^{\frac{ia}{\omega} \sin
(\omega t)} \end{array} \right). \label{matv} \eea

Using Eq.~\eqref{hf2} and the identity
$e^{iz \sin \phi} = \sum_{n=-\infty}^\infty J_n (z) e^{in \phi}$,
where the Bessel functions satisfy $J_n (-z) = J_{-n} (z) = (-1)^n
J_n (z)$, ~\cite{abramowitz} we find that the Floquet Hamiltonian
within sector 1 is \beq H_F ~=~ \left( \begin{array}{cc}
-1 & 1 \\
1 & -1 \end{array} \right) ~2 g^2 ~\sum_{n=-\infty}^\infty
\frac{[J_n (a/\omega)]^2}{U ~+~ n \hbar \omega}. \label{mat} \eeq We
see that one of the eigenstates of $H_F$ is the state $(|1 \rangle +
| 2 \rangle)/\sqrt{2}$ with eigenvalue zero (hence Floquet
eigenvalue equal to 1); this is one of the three spin-triplet
states, the other two being the ones given in Eq.~\eqref{2states}.
The other eigenstate of $H_F$ is $(|1 \rangle - | 2
\rangle)/\sqrt{2}$ which is a spin-singlet state, and the eigenvalue
is $-4 g^2 \sum_{n=-\infty}^\infty [J_n (a/\omega)]^2 /(U + n \hbar
\omega)$. Hence, in the spin language, the Floquet Hamiltonian has
the form \beq H_F ~=~ 4 g^2 ~\sum_{n=-\infty}^\infty \frac{[J_n
(a/\omega)]^2}{U ~+~ n \hbar \omega} ~\left( \frac{{\vec S}_1 \cdot
{\vec S}_2}{\hbar^2} ~-~ \frac{1}{4}\right). \label{ham6} \eeq

We have so far discussed some ways of calculating the Floquet
Hamiltonian perturbatively. We will now show that the Floquet
unitary can also be calculated perturbatively~\cite{bilitewski15}.
Given a time-dependent Hamiltonian $H(t)$ (which may not commute
with itself at different times), we define a time-evolution operator
as \beq U(t,0) ~=~ {\cal T} e^{-(i/\hbar) \int_{0}^{t} dt H (t)}.
\label{ut1} \eeq Now suppose
that $H(t) = H_0 (t) + V$, where $H_0$ is time-dependent but
exactly solvable, and $V$ is a time-independent term which we want to treat
perturbatively. We denote the time-evolution operator corresponding
to $H_0$ as $U_0 (t_2,t_1)$, so that \beq i \hbar \frac{\partial U_0
(t,0)}{\partial t} ~=~ H_0 (t) U_0 (t,0). \label{sch6} \eeq

Next, we define states in the interaction picture as \beq \psi^I (t)
~=~ U_0 (0,t) \psi (t). \label{psiI} \eeq This satisfies the
Schr\"odinger equation \beq i \hbar \frac{\partial \psi^I}{\partial
t} ~=~ V^I (t) \psi^I (t), \label{sch7} \eeq where \beq V^I (t) ~=~
U_0 (0,t) V U_0 (t,0). \label{vI} \eeq The corresponding
time-evolution operator \beq U^I (t,0) ~=~ {\cal T} e^{-(i/\hbar)
\int_0^t dt' V^I (t')}, \label{uI1} \eeq satisfies the equation \beq
i \hbar \frac{\partial U^I (t,0)}{\partial t} ~=~ V^I (t) U^I (t,0).
\label{sch8} \eeq Assuming the initial condition $U^I (0,0) = I$,
the solution of Eq.~\eqref{sch8} is \beq U^I (t,0) ~=~ I ~-~
\frac{i}{\hbar} ~\int_0^t dt' V^I (t') U^I (t',0). \label{uI2} \eeq
This provides an iterative way of calculating $U^I (t,0)$ in powers
of $V^I$. Thus one can write \bea U^I (t,0) &=&
I ~+~ \left( \frac{-i}{\hbar} \right) ~\int_0^t dt' V^I (t') \nonumber\\
&& +~\left(\frac{-i}{\hbar}\right)^2 \int_0^t dt_1 V^I(t_1)
\int_0^{t_1} dt_2 V^I(t_2) + ... . \nonumber\\
&=& I + U_1^I (T,0) + U_2^I(t,0) + ..., \label{uI3} \eea where the
ellipsis denotes higher order terms. Finally, the full
time-evolution operator is given by \beq U(t,0) ~=~ U_0 (t,0) U^I
(t,0). \label{ut2} \eeq In the case where $U_0(T,0)=I$, the Floquet
operator is obtained by setting $t=T$ in Eq.~\eqref{ut2} and is given by
\begin{eqnarray} H_F &=& \frac{i\hbar }{T} \left[U_1^I(T,0) + (U_2^I(T,0)-
(U_1^I(T,0))^2/2) + \cdots \right]. \nonumber\\ \label{flham1} \end{eqnarray}

We end this section with a few comments regarding the FPT technique
which deals directly with $U$. First, we note that the perturbation
involving $U$ is useful if one attempts to compute higher order
terms since it provides a straightforward and systematic way of
obtaining such terms especially when $U_0(T,0)=I$. This method has
indeed been used to compute second and third order perturbative
terms in several interacting many-body systems which is otherwise
difficult \cite{roopayan20,fr3}. Second, in contrast to the wave
function method, the truncation of the perturbation series
necessarily leads to loss of unitarity of $U$. In the case when
$U_0(T,0)=I$, this can be remedied by exponentiating the terms in
$U^I(T,0)$. In contrast, such unitarization procedure is neither
unique nor straightforward if $U_0(T,0) \ne 0$. However, sometimes
special dynamical symmetries of $U(T,0)$ may help one to carry out
the task \cite{fcft3}.

\section{Other Methods}
\label{sec:om}

In this section, we present a brief discussion of three other
methods which have been used in the literature to compute the Floquet
Hamiltonian of a driven system.

\subsection{Rotating wave approximation}
\label{ssec:rwa}

The rotating wave approximation (RWA) provides a way to calculate an
effective Hamiltonian by transforming to a `rotating frame' in such
a way that the Hamiltonian in this frame does not have any
time-dependent terms to lowest order~\cite{rev1,eckardt17}. To see
how this works, consider a time-dependent Hamiltonian $H(t)$ and a
general wave function $\psi (t)$ satisfying the Schr\"odinger
equation $i \hbar \partial \psi /\partial t = H \psi$. Given a
unitary operator $W(t)$ which transforms to a rotating frame, we
define a wave function in that frame as \beq \psi_R (t) ~=~ W(t)
\psi (t). \label{psir} \eeq We find that $\psi_R$ satisfies the
Schr\"odinger equation $i \hbar \partial \psi_R /\partial t = H_R
\psi_R$, where \beq H_R ~=~ W H W^{-1} ~+~ i \hbar \frac{\partial
W}{\partial t} W^{-1} \label{hr1} \eeq is the Hamiltonian in the
rotating frame.

Now suppose that $H(t) = H_0 (t) + V$, where $H_0 (t+T) = H_0 (t)$.
We can then try to choose $W (t)$ in such a way that the
time-dependent part of $H_R$ in Eq.~\eqref{hr1} is as small as
possible. For instance, if $H_0$ is much larger than $V$, and $H_0
(t)$ commutes with itself at different times, we choose \beq W(t)
~=~ e^{(i/\hbar) \int_0^t dt' H_0 (t')}. \label{wt1} \eeq Since $W$
and $H_0$ commute with each other at different times, we see from
Eqs.~(\ref{hr1}-\ref{wt1}) that \beq H_R ~=~ W V W^{-1}, \label{hr2}
\eeq which implies that the large term $H_0$ has disappeared in
going from $H$ to $H_R$.

We would like $W$ to satisfy $W(t+T) = W(t)$. Eq.~\eqref{wt1}
implies that this will be true if $H_0 (t)$ has a complete set of
orthonormal eigenstates $|n \rangle$ with eigenvalue $E_n (t)$, such
that \beq \int_0^T dt E_n (t) ~=~ 0 \label{inten} \eeq for all $n$;
this turns out to be true in many problems. Then $H_R$ in
Eq.~\eqref{hr2} will satisfy $H_R (t+T) = H_R (t)$.
Eq.~\eqref{inten} also implies Eq.~\eqref{res1} which means that we
have to do degenerate FPT.

Next, inserting the identity operator, $I = \sum_n | n \rangle
\langle n |$, on the left and right sides of $H_R$ in
Eq.~\eqref{hr2}, we obtain \beq H_R (t) ~=~ \sum_{m,n} |m \rangle
\langle m | V | n \rangle \langle n | e^{(i/\hbar) \int_0^t dt' [E_m
(t') - E_n (t')]}. \label{hr3} \eeq If we now do a FM expansion
with $H_R$, the first term is
\bea H_M^{(0)} &=& \frac{1}{T} \int_0^T ~dt H_R (t) \nonumber \\
&=& \sum_{m,n} |m \rangle \langle n | ~\frac{\langle m | V | n
\rangle}{T} \nonumber \\
&& \times ~\int_0^T dt e^{(i/\hbar) \int_0^t dt' [E_m (t') - E_n
(t')]}. \label{hm0} \eea The matrix elements of $H_M^{(0)}$ in
Eq.~\eqref{hm0} agree with those of the Floquet Hamiltonian $H_F$
given in Eq.~\eqref{hmn1}. (Note that the second line of
Eq.~\eqref{hmn1} vanishes due to the condition in
Eq.~\eqref{inten}). We thus see that there is a connection between
FPT and RWA if the periodically driven term is the one with the
largest coefficient. However, the higher order terms that we get in
the FM expansion of $H_R (t)$ have no counterparts in $H_F$
obtained from FPT.

A simple use of the RWA is to solve the problem of a spin-1/2
particle in a magnetic field which is rotating about one
axis~\cite{rev1}. We consider the Hamiltonian \beq H ~=~ B_1
\sigma^z ~+~ B_2 ~[\cos (\omega t) \sigma^x ~+~ \sin (\omega t)
\sigma^y]. \label{ham7} \eeq Using Eq.~\eqref{hr1}, we find that the
operator \beq W ~=~ e^{i \omega t/2} ~e^{i (\omega t/2) \sigma^z}
\label{w1} \eeq which rotates by an angle $\omega t$ around the
$\hat z$-axis gives the Hamiltonian \beq H_R ~=~ (B_1 ~-~
\frac{\hbar \omega}{2}) \sigma^z ~+~ B_2 \sigma^x ~-~ \frac{\hbar
\omega}{2} I. \label{hr4} \eeq (The factor of $e^{i \omega t/2}$ has
been included in Eq.~\eqref{w1} to ensure that $W (t+T) = W (t)$).
We see that $H_R$ is completely time-independent and has the
eigenvalues \beq E_{\pm} ~=~ -~ \frac{\hbar \omega}{2} ~\pm~
\sqrt{\left( B_1 ~-~ \frac{\hbar \omega}{2} \right)^2 ~+~ B_2^2}.
\label{epm1} \eeq We can use the eigenvalues $E_\pm$ and the
corresponding orthonormal eigenstates $\psi_\pm$ of $H_R$ to find
the general solution for $\psi (t)$ through Eq.~\eqref{psir}.

No assumptions were made about the relative magnitudes of $\hbar
\omega$, $B_1$ and $B_2$ while deriving Eq.~\eqref{hr4}. We now note
that when $\omega \to \infty$, $H_R$ diverges instead of approaching
the time-averaged value $(1/T) \int_0^T dt H = B_1 \sigma^z$ which
is finite. This can be fixed as follows. An examination of
Eq.~\eqref{epm1} shows that if $\omega \to \infty$, $E_+ \to - B_1$
and $E_- \to B_1 - \hbar \omega$, where we have ignored terms of
order $1/\omega$; in the same limit, $\psi_+ \to (0, 1)$ and $\psi_-
\to (1, 0)$. Since the Floquet eigenvalues $e^{-i E_\pm T/\hbar}$
remain invariant if the quasienergies $E_\pm$ are shifted by
arbitrary integer multiples of $\hbar \omega$, we use this freedom
to add $\hbar \omega$ to $E_-$ while keeping $E_+$ as it is. This
gives us new quasienergies \beq E'_{\pm} ~=~ \pm \left[ ~-~
\frac{\hbar \omega}{2} ~+~ \sqrt{\left( B_1 ~-~ \frac{\hbar
\omega}{2} \right)^2 ~+~ B_2^2} \right] \label{epm2} \eeq which tend
to $\mp B_1$ as $\omega \to \infty$. Combining these with the
eigenstates $\psi_\pm$ of $H_R$ in Eq.~\eqref{hr4}, we construct a
new Hamiltonian $H'_R = E'_+ ~\psi_+ \psi_+^\dagger + E'_- ~\psi_-
\psi_-^\dagger$ which gives \bea H'_R &=& \frac{-~ \frac{\hbar
\omega}{2} ~+~ \sqrt{\left( B_1 ~-~ \frac{\hbar \omega}{2} \right)^2
~+~ B_2^2}}{\sqrt{\left(
B_1 ~-~ \frac{\hbar \omega}{2} \right)^2 ~+~ B_2^2}} \nonumber \\
&& \times~ \left[(B_1 ~-~ \frac{\hbar \omega}{2}) \sigma^z ~+~ B_2
\sigma^x \right]. \label{hr5} \eea In the limit $\omega \to \infty$,
$H'_R \to B_1 \sigma^z$ as desired.

We now present another application of the RWA. We consider a
tight-binding model of spinless particles (which may be either
fermions or bosons) in one dimension placed in an electric field
${\cal E} (t)$ which varies in time with a period
$T$~\cite{eckardt17}. If the spacing between neighboring sites is
$d$, the electric field can be put in as an on-site potential $- q n
d {\cal E} (t)$ at site $n$, where $q$ is the charge of the
particle. The complete Hamiltonian is \beq H ~=~
\sum_{n=-\infty}^\infty [-g (c_n^\dagger c_{n+1} + c_{n+1}^\dagger
c_n) - q d {\cal E} (t) n c_n^\dagger c_n]. \label{ham8} \eeq We can
now eliminate the on-site potential in Eq.~\eqref{ham8} and move the
time-dependence to the phases of the nearest-neighbor hoppings by
transforming with \beq W (t) ~=~ e^{-(i q d/\hbar) \int_0^t dt'
{\cal E} (t') \sum_{n=-\infty}^\infty n c_n^\dagger c_n}.
\label{wt2} \eeq Using Eq.~\eqref{hr1}, we find that the Hamiltonian
in the rotating frame is \bea H_R &=& - ~g ~\sum_{n=-\infty}^\infty
[e^{(i q d/\hbar) \int_0^t dt'
{\cal E} (t')} c_n^\dagger c_{n+1} \nonumber \\
&& ~~~~~~~~~~~~~+~ e^{-(i q d/\hbar) \int_0^t dt' {\cal E} (t')}
c_{n+1}^\dagger c_n]. \label{hr6} \eea (We note that the
transformation from Eq.~\eqref{ham8} to \eqref{hr6} is a gauge
transformation which takes us from an electrostatic potential
defined at a site $n$ to a vector potential which appears in the
phase of the hopping between sites $n$ and $n+1$ according to the
Peierls prescription). Next, we define \beq f(t) ~=~
\frac{qd}{\hbar} ~\int_0^t dt' {\cal E} (t'). \label{ft} \eeq Let us
assume that $\int_0^T dt {\cal E} (t) = 0$; then the periodicity of
${\cal E} (t)$ also implies the periodicity of $f(t)$ and therefore
of $e^{i f(t)}$. We then find, by going to momentum space, that
Eq.~\eqref{hr6} takes the form \beq H_R ~=~ \int_{-\pi/d}^{\pi/d}
\frac{dk}{2\pi /d} ~[-2g \cos (k + f(t))]~ c_k^\dagger c_k.
\label{hr7} \eeq

Now suppose that $e^{if}$ has the Fourier expansion \beq e^{i f(t)}
~=~ \sum_{m=-\infty}^\infty ~F_m ~e^{-im \omega t}, \label{fft} \eeq
so that $(1/T) \int_0^T dt e^{i f(t)} = F_0$. If we now do a FM
expansion of the Hamiltonian in Eq.~\eqref{hr7}, we find that only
the first term given by $H_M = (1/T) \int_0^T dt H_R$ survives, and
we obtain \bea H_M &=& \int_{-\pi/d}^{\pi/d} \frac{dk}{2\pi /d} ~E_k
~c_k^\dagger c_k,
\nonumber \\
E_k &=& - ~g ~(F_0 e^{ik} ~+~ F_0^* e^{-ik}). \label{hf3} \eea As an
example, if $(q d/\hbar) {\cal E} (t) = a \cos (\omega t)$, we have
$F_0 = J_0 (a/\omega)$. The energy-momentum dispersion is then given
by $E_k = -2 g J_0 (a/\omega) \cos k$, to be compared with the
dispersion $- 2 g \cos k$ for the undriven system. Note that a flat
band is generated if $a$ and $\omega$ satisfy $J_0 (a/\omega) = 0$,
and this leads to the dynamical localization of any wave packet
since the group velocity $dE_k/dk = 0$ for all values of $k$.

\subsection{Floquet-Magnus resummation}
\label{ssec:fmr}

For periodically kicked models where the Floquet unitary
is of the form (assuming $\hbar=1$)
\beq U(T)=\exp(-iTH)\exp(-i \epsilon V)=U_0 \exp(-i \epsilon V),
\label{replica1} \eeq where $[H,V] \neq 0$ and $V$ represents a
periodic kicking term (e.g., see Eq.~\eqref{fermion2}) with $\epsilon
\ll 1$ being a small parameter, a replica trick can be
used~\cite{BCHreplicaPRL} to write the Floquet Hamiltonian as a
power series in $\epsilon$ instead of using the standard BCH formula
(Eq.~\eqref{BCH}) which yields a power series in $T$. Closed-form
expressions for $H_F$ can then be computed in some
cases~\cite{BCHreplicaPRL,bukov1,bukov2}. Using the replica trick,
\begin{eqnarray} TH_F &=& i\ln U(T) \nonumber \\
&=& i \lim_{\rho \to 0} \frac{1}{\rho} \left( U^\rho(T)-1 \right).
\label{replica2} \end{eqnarray}
Using a Taylor expansion in $\epsilon$ then yields the expression
\beq i\ln U(T)=i\lim_{\rho \to 0} \frac{1}{\rho} \left(\sum_{r=0}^\infty
\left[\frac{1}{r!} \left(\partial_\epsilon^r U^\rho(T) \right)
\vert_{\epsilon=0}\right] \epsilon^r \right). \label{replica3} \eeq
Interchanging the sum and limit in Eq.~\eqref{replica3} then gives the
power series
\begin{eqnarray} H_F &=& \frac{1}{T} \sum_{r=0}^\infty \Gamma_r \epsilon^r,
\mbox{~~~}\mathrm{where} \nonumber \\
\Gamma_r &=& \frac{i}{r!} \lim_{\rho \to 0} \frac{1}{\rho}\left[ \left(
\partial_\epsilon^r U^\rho(T) \right)\vert_{\epsilon=0}\right].
\label{replica4} \end{eqnarray}
So far the discussion was general, and further progress in computing
$\Gamma_r$ is made by restricting to $U(T)$ of the form shown in
Eq.~\eqref{replica1}. The general expression for $\Gamma_r$ is worked
out in Ref.~\onlinecite{BCHreplicaPRL}. We restrict ourselves to the
$\mathrm{O}(\epsilon)$ term which only requires $\Gamma_1$ that can
be obtained straightforwardly (assuming an integer replica index
$\rho$): \beq \Gamma_1 = i \lim_{\rho \to 0} \frac{U_0^\rho}{\rho}
\left[\sum_{m=0}^{\rho-1}U_0^{-m}VU_0^m \right]. \label{replica5}
\eeq The limit $\rho \to 0$ is then taken following an analytic
continuation to arbitrary real values. We illustrate this using an
example from Ref.~\onlinecite{bukov2} where in Eq.~\eqref{replica1},
we take a one-dimensional driven Ising model with
\begin{eqnarray}
H &=& \sum_j (J \sigma^z_j \sigma^z_{j+1} + h_z \sigma_j^z),
\label{replica6} \\
V(t) &=& \sum_{n=-\infty}^{\infty} V \delta (t-nT), \quad V= h_x
\sum_j \sigma_j^x. \nonumber \end{eqnarray} Since $H$ contains a sum
of commuting terms, $U_0^m$ can be evaluated to finally give
\begin{widetext}
\begin{eqnarray}
U_0^{-m} V U_0^m &=& - ~\frac{h_x}{2} \sum_j \sin(4mJT) \left[ \frac{}{}
\sigma_{j-1}^z (\cos(2mh_zT)\sigma_j^y+ \sin(2mh_zT) \sigma_j^x) \right .
\nonumber \\
&& ~~~~~~~~~~~~~~~~~~~~~~~~~~~~~+ \left.(\cos(2mh_zT)\sigma^y_j + \sin(2mh_zT)
\sigma_j^x) \sigma_{j+1}^z \frac{}{}\right] \nonumber \\
&& - ~h_x \sum_j \sin^2(2mJT) ~\sigma_{j-1}^z \left[\frac{}{}\cos(2mh_zT)
\sigma_j^x-\sin(2mh_zT)\sigma_j^y\right]\sigma^z_{j+1}\nonumber \\
&& + ~h_x \sum_j \cos^2(2mJT) \left[\frac{}{}\cos(2mh_zT)\sigma^x_j -
\sin(2mh_zT)\sigma_j^y \right]. \label{replica7} \end{eqnarray}
\end{widetext}
To evaluate the sum over $m$ in Eq.~\eqref{replica5}, we use a mode
expansion for $\sin(\theta)=(\exp(i\theta)-\exp(-i\theta))/(2i)$ and
$\cos(\theta)=(\exp(i\theta)+\exp(-i\theta))/2$ in
Eq.~\eqref{replica7}, collect terms with different
exponents, and use summations of geometric sequences to finally get
\begin{widetext}
\begin{eqnarray}
\sum_{m=0}^{\rho-1} U_0^{-m}VU_0^m &=& - ~h_x \left[ \frac{1}{8i}
\left(F^{-}_{4JT-2h_zT}(\rho)+F^{-}_{4JT+2h_zT}(\rho) \right) \sum_j
(\sigma_j^z \sigma_{j+1}^y+\sigma_j^y \sigma_{j+1}^z) \right. \nonumber \\
&& ~~~~~~~~+ \left. \frac{1}{8} \left(F^{+}_{4JT-2h_zT}(\rho)-
F^{+}_{4JT+2h_zT}(\rho) \right) \sum_j (\sigma_j^x \sigma_{j+1}^z+\sigma_j^z
\sigma_{j+1}^x) \right. \nonumber \\
&& ~~~~~~~~- \left. \frac{1}{8} \left(F^{+}_{4JT-2h_zT}(\rho)+
F^{+}_{4JT+2h_zT}(\rho) -2F^{+}_{-2h_Z T}(\rho)\right) \sum_j
(\sigma_{j-1}^z \sigma_j^x \sigma_{j+1}^z)\right. \nonumber \\
&& ~~~~~~~~+ \left. \frac{i}{8} \left(F^{-}_{4JT-2h_zT}(\rho)-
F^{-}_{4JT+2h_zT}(\rho) -2F^{-}_{-2h_Z T}(\rho)\right) \sum_j
(\sigma_{j-1}^z \sigma_j^y \sigma_{j+1}^z)\right. \nonumber \\
&& ~~~~~~~~- \left. \frac{1}{8} \left(F^{+}_{4JT-2h_zT}(\rho)-
F^{+}_{4JT+2h_zT}(\rho) +2F^{+}_{-2h_Z T}(\rho)\right) \sum_j
(\sigma_j^x)\right. \nonumber \\
&& ~~~~~~~~+ \left. \frac{i}{8} \left(F^{-}_{4JT-2h_zT}(\rho)-
F^{-}_{4JT+2h_zT}(\rho) +2F^{-}_{-2h_Z T}(\rho)\right) \sum_j
(\sigma_j^y)\right], \label{replica8} \end{eqnarray}
\end{widetext}
where
\begin{eqnarray}
F^-_\chi(\rho) &=& -i \frac{\cos(\chi \rho -\chi/2)-\cos(\chi/2)}{\sin(
\chi/2)}, \nonumber \\
F^+_\chi(\rho) &=& \frac{\sin(\chi \rho -\chi/2)+\sin(\chi/2)}{\sin(\chi/2)}.
\label{replica9} \end{eqnarray}

Finally using Eq.~\eqref{replica5} and taking the limit $\rho \to 0$,
we get the expression for $H_F$ to $\mathrm{O}(\epsilon)$,
\begin{widetext}
\begin{eqnarray}
H_F &=& \sum_j (J \sigma_j^z \sigma_{j+1}^z +h_z \sigma^z_j)+\epsilon h_x
\left[ \frac{}{} J \sum_j (\sigma^z_j \sigma^y_{j+1} + \sigma^y_j
\sigma^z_{j+1})\right. \nonumber \\
&& ~~~~~~~+ \frac{1}{4} \left. [(2J+h_z)\cot(2JT+h_zT)-(2J-h_z)
\cot(2JT-h_zT)]\sum_j (\sigma^x_j \sigma^z_{j+1} + \sigma^z_j
\sigma^x_{j+1}) \right. \nonumber \\
&& ~~~~~~~+ \frac{1}{4} \left. [(2J-h_z)\cot(2JT-h_zT)-(2J+h_z)
\cot(2JT+h_zT)-2h_z \cot(h_zT)]\sum_j (\sigma^z_{j-1} \sigma^x_{j}
\sigma^z_{j+1}) \right. \nonumber \\
&& ~~~~~~~+ \frac{1}{4} \left. [(2J-h_z)\cot(2JT-h_zT)+(2J+h_z)
\cot(2JT+h_zT)+2h_z \cot(h_zT)]\sum_j \sigma^x_{j} \right. \nonumber \\
&& ~~~~~~~+ \left. h_z \sum_j \sigma_j^y \frac{}{}\right]. \end{eqnarray}
\end{widetext}
In Ref.~\onlinecite{BCHreplicaPRL}, it was demonstrated for one-dimensional
kicked Ising models that long prethermal Floquet regimes can exist in the
thermodynamic limit even at intermediate and low drive frequencies that
are governed by $H_F$ obtained from the replica expansion.

\subsection{Hamiltonian Flow Method}
\label{ssec:hf}

The Hamiltonian flow method was initially applied to
time-independent Hamiltonians by Wegner \cite{wegner94, kehrein06}.
The idea behind this method is to obtain an effective Hamiltonian,
via generation of a flow, which is diagonal in a chosen basis
(usually chosen to be the non-interacting single particle
eigenstates). The flow is designed such that the off-diagonal terms
reduce gradually as the flow continues. Such a flow is characterized
by a parameter $\ell$ and is implemented via an anti-unitary
transformation $\eta(\ell)$. The flow equation, in terms of this
anti-unitary operator $\eta$ is given by \cite{wegner94}
\begin{eqnarray} \frac{d H(\ell)}{d\ell} &=& \left[ \eta(\ell), H(\ell)
\right]. \label{flowham1} \end{eqnarray} Of course, the key issue
here is the choice of $\eta(\ell)$ and this depends on the system at
hand. For example, for a two-level system with $H(\ell)\equiv H_t$
given by
\begin{eqnarray} H_t &=& \sum_{i=1,2} \epsilon_i |i\rangle + \left(\Delta_{12}
|1\rangle \langle 2| + {\rm H.c.} \right), \label{flow2level} \end{eqnarray}
it was shown that one can choose $\eta =- [H_t,H_D]$, where $H_D$ is
the diagonal part of $H_t$. This choice leads to (taking $\Delta_{12}$ to
be real for simplicity)
\begin{eqnarray} \frac{\partial \Delta_{12}}{\partial \ell} &=& [\eta, H]_{12}
\nonumber\\
&=&\sum_{j=1,2} (\epsilon_1+ \epsilon_2 - 2 \epsilon_j)
\Delta_{1j}\Delta_{j 2}. \label{floweq1} \end{eqnarray} where
$\Delta_{11} = \epsilon_1$ and $\Delta_{22} = \epsilon_2$. It was
shown in Ref.\ \onlinecite{wegner94} that these equations lead to
\begin{eqnarray} \Delta_{12}(\ell) = \Delta_{12}(\ell=0)
\exp[-(\epsilon_1-\epsilon_2)^2 \ell]. \label{flowres1} \end{eqnarray}
Thus the Hamiltonian reduces to a diagonal form with increasing
$\ell$. The method gets more complicated with increasing number of energy
levels, and in most cases the flow equations require a numerical solution.

As shown by several authors \cite{flflow1,fiete18}, this method can
be modified to be applied to periodically driven systems. Here one
tries to generate such a flow to compute the Floquet Hamiltonian. To
this end, the Hamiltonian is divided into two parts, $H_0$ and $V$.
Typically, $H_0$ is chosen to be the first term in the FM expansion
of $H_F$ so that
\begin{eqnarray} H_0 &=& \frac{1}{T} \int_0^T dt H(t), \quad V= H(t)-H_0.
\label{hampart} \end{eqnarray}
Note that $V \to 0$ for $T \to 0$, so that the method, for such a
choice of the division of $H(t)$, is naturally accurate at high drive
frequency. Next, the flow parameter $\ell$ and $\eta$ are introduced through
the transformation $U = \exp(\eta[\ell,t])$, where
\begin{eqnarray} \eta[\ell,t] &=& -i \frac{\ell}{T} \int_0^t dt_1 V[\ell,t_1],
\label{etaldef} \end{eqnarray}
where we have extended the definition of $H(t)$ to $H[\ell,t]$ with
\begin{eqnarray}
V[\ell,t] &=& \frac{1}{T} \int_0^T dt_1 ~(H[\ell,t]-H[\ell,t_1]).
\label{vtrans} \end{eqnarray}
Note that $V[0,t]= V(t)$ and our aim is show that $V[\infty,t]=0$.

To obtain the flow equation, one generates an infinitesimal
change by $\delta \ell$ and relates $H[\ell+\delta \ell,t]$ to
$H[\ell,t]$. A straightforward calculation charted out in Ref.\
\onlinecite{fiete18} shows that this leads to the flow equation
\begin{eqnarray} \frac{d H[\ell,t]}{d \ell} &=& - V[\ell,t] + i \int_0^t
dt_1[V[\ell,t],H[\ell,t]]. \label{fleq} \end{eqnarray}
The general nature of the flow from this equation is quite
complicated. However, consideration simplification occurs at high
frequencies where the second term can be neglected. In this case the
equation reduces to $dV[\ell,t]/d\ell= -V[\ell,t]$ and leads to $V=
V(t) e^{-\ell}$. Thus $ V[\infty,t] =0$ and $H= H_0$ (Eq.\
\eqref{hampart}) which is the standard FM result.

The existence of such fixed points of the flow equation at lower drive
frequencies is not simple to prove. This issue and the application
of the method to a simple spin model where the flow equations can be
written down in a straightforward manner have been studied in Ref.\
\onlinecite{fiete18}. However, the applicability of this method to
more complicated interacting many-body systems and the fate of the
flow equation at lower frequency remain interesting open problems.

\section{Discussion}
\label{sec:diss}

In this review, we have aimed to provide a pedagogical discussion
of several analytic methods for computing the Floquet Hamiltonian
for interacting many-body systems. Out of these, the Floquet-Magnus
expansion method, the adiabatic-impulse approximation and the
Floquet perturbation theory are treated in detail. Three other
methods, namely, the rotating wave approximation, the Floquet-Magnus
resummation method, and the Hamiltonian flow technique have been
discussed briefly.

The Floquet-Magnus expansion is probably the most widely used
perturbative method in the literature. This technique treats inverse
of the drive frequency (in units of $J_{\mathrm{loc}}/\hbar$, where
$J_{\mathrm{loc}}$ is a typical energy scale of the system) as the
perturbation parameter and is therefore expected to be accurate in
the high drive frequency regime. Moreover, it provides a
perturbative expansion which maintains unitarity of $U$ at each
order in perturbation theory. Also, the method has been quite
successful in providing a qualitatively accurate picture of
drive-induced generation of topologically non-trivial Floquet states
and the presence of a long prethermal timescale in interacting
many-body systems at high drive frequencies. The main weakness of
this method is two-fold. First, the radius of convergence of the
perturbation expansion and its regime of validity is difficult to
determine. Second, the method may lead to qualitatively wrong
pictures at intermediate and low drive frequencies
\cite{roopayan20}. The resolution of these issues which might
provide one with a more complete picture of the Floquet-Magnus
expansion method is a long standing challenge; some progress in this
direction has been made recently \cite{pretherm1, BCHreplicaPRL}.

In contrast, the adiabatic-impulse approximation leads to a
method which is more accurate at low drive frequencies.
This makes the method ideal for studying properties of driven systems in
regimes where most other approximation methods fail. Moreover, the
method provides access to the micromotion of the system; thus it
allows one to obtain information about the phase bands which carries
significantly more information about the properties of the driven
system than the Floquet Hamiltonian. The key disadvantage of the
method is that its application is limited to a class of integrable
models; for more complicated non-integrable modes, it gets extremely
complicated. The possibility of generalizing this method so
that it may be applied to non-integrable and/or multi-band models is
an open problem.

The Floquet perturbation theory, in contrast to both the methods
discussed above, does not treat the drive frequency or it's inverse
as a perturbation parameter. The perturbation parameter for this
method is the ratio of amplitudes of the terms in the Hamiltonian;
the largest term in the Hamiltonian (which consists of commuting
terms) is treated exactly and the contribution of the rest, smaller,
terms are assessed using perturbation theory. The method provides
qualitatively accurate results even when the drive frequency is
small compared to the amplitude of the largest term and can
therefore access regimes where the Floquet-Magnus expansion fails.
When this largest amplitude term is also the term which implements
the periodic drive, this method is analogous to the rotating wave
approximation to leading order. However, for other cases, the
Floquet perturbation theory leads to different, more accurate,
results compared to the rotating wave approximation
\cite{roopayan20}. Moreover, the first order contribution to the
Floquet Hamiltonian in this method already constitutes a resummation
of an infinite class of terms in the Floquet-Magnus expansion; this
has been explicitly demonstrated in the context of ultracold Rydberg
atoms subjected to a square pulse protocol in Ref.\
\onlinecite{flscar1}. The key disadvantages of this method are
two-fold. First, it does not automatically lead to an unitary
evolution operator at a given order in perturbation theory and a
separate unitarization procedure is necessary. This procedure need
not be unique specially when the zeroth order Floquet Hamiltonian,
$H_F^{(0)}$, does not vanish. Second, the method is difficult to
apply if the largest amplitude term in the Hamiltonian is
complicated so that its contribution to $U$ cannot be determined
exactly. These issues constitute some open problems relevant for FPT.

Next, we note that we have not provided a discussion of numerical
methods for computing $U$ or $H_F$. This, in our opinion, warrants a
separate review. Here we briefly comment that the standard method
for this involves exact diagonalization (ED), specially when one
wants an access to all states in the Hilbert space of the driven
model necessary for tracking the dynamics at long times
\cite{edref1}. For this one proceeds as follows. First, one
decomposes $U(T,0)$ into a product of $N$ Trotter steps ($U_j =
U(t_{j-1} +\delta t, t_{j-1})$ being the unitary at the $j$-th step
and $\delta t=T/N$), where the value of $N$ depends on both the
nature of the Hamiltonian and the drive protocol used. Here we note
that $N$ is of order 1 if we use discrete protocols such as a square
pulse or periodic kicks since for these protocols $H(T)$ remains
constant for a large part of the drive cycle. In contrast, for
continuous drives $N \gg 1$ and therefore such drive protocols are
difficult to treat numerically. Next, one expresses $U_j$ using the
eigenenergies and eigenvectors of $H_j \equiv H(t_{j-1} +\delta
t/2)$ which can be obtained using ED. Finally one computes the
product over $U_j$ to construct $U$ and diagonalizes to obtain the
Floquet eigenvectors and quasienergies. Clearly, the numerical cost
of this method depends both on the Hilbert space dimension of the
many-body system and $N$; thus this method is most useful for
one-dimensional systems driven by discrete protocols. There are
several other numerical methods for computing the Floquet spectrum
of a driven many-body system \cite{numref1}, but we will not address
them here.

We also note that quasiperiodically-driven many-body quantum
systems, with two or more incommensurate drive frequencies, have
received attention only
recently~\cite{quasi1,quasi2,quasi3,quasi4,quasi5,quasi6,quasi7,quasi8,quasi9}.
While the steady state is again expected to be described by a
featureless infinite temperature ensemble, much less is known about
possible prethermal phases in such settings. In case such prethermal
phases exist (see Refs.~\onlinecite{quasi7} and
\onlinecite{quasi9}), they would likely constitute a much richer
class than their Floquet analogs. Finding reliable perturbative
approaches, particularly beyond the high-frequency regime where a
generalization of the Floquet-Magnus expansion to quasiperiodic
drives exists~\cite{quasiFM}, remains an uncharted territory.

Finally, we would like to point out that in this review we have not
discussed path integrals methods for periodically driven quantum
systems. It is well known that study of non-equilibrium dynamics
using path integrals usually requires use of the Schwinger-Keldysh
formalism \cite{kelref1}. This method is used for study of open
systems \cite{openref}, driven superconductors \cite{kelsup}, and
systems with dissipation treated within Lindblad approach
\cite{linref}. It is also to be noted for closed integrable systems,
path integral technique may be used for computation of the Floquet
Hamiltonian of a periodically driven system for certain protocols
without resorting to Keldysh techniques \cite{nicolas1}. Overall,
the path integral method is usually more widely used for treatment
of open quantum systems and we do not study its details in this review.

In conclusion we have compared and contrasted several analytic,
albeit perturbative, techniques for computation of a periodically
driven many-body system. Our review, while not entirely exhaustive,
provides a pedagogical introduction to the technical details of
several such methods and addresses some of the applications of these
methods to a number of recent problems in periodically driven
quantum many-body systems.

\section{Acknowledgments}

The work of A.S. is partly supported through the Max Planck Partner Group
program between the Indian Association for the Cultivation of Science (Kolkata)
and the Max Planck Institute for the Physics of Complex Systems (Dresden).
D.S. thanks DST, India for Project No. SR/S2/JCB-44/2010 for financial support.

\end{document}